\documentstyle[12pt]{article}

\catcode`\@=11
\def\marginnote#1{}
\newcount\hour
\newcount\minute
\newtoks\amorpm
\hour=\time\divide\hour by60 \minute=\time{\multiply\hour by60
\global\advance\minute by-\hour}\edef\standardtime{{\ifnum\hour<12
\global\amorpm={am}%
        \else\global\amorpm={pm}\advance\hour by-12 \fi
        \ifnum\hour=0 \hour=12 \fi
        \number\hour:\ifnum\minute<10
0\fi\number\minute\the\amorpm}}
\edef\militarytime{\number\hour:\ifnum\minute<10
0\fi\number\minute}

\def\draftlabel#1{{\@bsphack\if@filesw {\let\thepage\relax
   \xdef\@gtempa{\write\@auxout{\string
      \newlabel{#1}{{\@currentlabel}{\thepage}}}}}\@gtempa
   \if@nobreak \ifvmode\nobreak\fi\fi\fi\@esphack}
        \gdef\@eqnlabel{#1}}
\def\@eqnlabel{}
\def\@vacuum{}
\def\draftmarginnote#1{\marginpar{\raggedright\scriptsize\tt#1}}
\def\draft{\oddsidemargin -.5truein
        \def\@oddfoot{\sl preliminary draft \hfil
        \rm\thepage\hfil\sl\today\quad\militarytime}
        \let\@evenfoot\@oddfoot \overfullrule 3pt
        \let\label=\draftlabel
        \let\marginnote=\draftmarginnote

\def\@eqnnum{(\theequation)\rlap{\kern\marginparsep\tt\@eqnlabel}%
\global\let\@eqnlabel\@vacuum}  }

\def\numberbysection{\@addtoreset{equation}{section}
        \def\theequation{\thesection.\arabic{equation}}}

\def\underline#1{\relax\ifmmode\@@underline#1\else
 $\@@underline{\hbox{#1}}$\relax\fi}

\catcode`@=12 \relax

\numberbysection

\topmargin by -\headsep

\def\sp{\partial\!\!\!/}
\def\sa{A\!\!\!/}

\def\br{\begin{eqnarray}}
\def\er{\end{eqnarray}}
\def\be{\begin{equation}}
\def\ee{\end{equation}}

\def\({\left(}
\def\){\right)}

\relax

\def\a{\alpha}

\def\b{\beta}

\def\d{\delta}
\def\D{\Delta}

\def\g{\gamma}

\def\l{\lambda}
\def\L{\Lambda}
\def\m{\mu}

\def\pa{\partial}

\def\s{\sigma}

\def\tp0{\Theta_{+}^{(0)}}
\def\tm0{\Theta_{-}^{(0)}}

\def\vp{\varphi}

\def\f#1#2#3 {f^{#1#2}_{#3}}

\def\win1{{\sf w_{1+\infty}}}

\def\Win1{{\sf W_{1+\infty}}}

\def\rlx{\relax\leavevmode}
\def\inbar{\vrule height1.5ex width.4pt depth0pt}
\def\IZ{\rlx\hbox{\sf Z\kern-.4em Z}}
\def\IR{\rlx\hbox{\rm I\kern-.18em R}}
\def\IC{\rlx\hbox{\,$\inbar\kern-.3em{\rm C}$}}
\def\IN{\rlx\hbox{\rm I\kern-.18em N}}
\def\IO{\rlx\hbox{\,$\inbar\kern-.3em{\rm O}$}}
\def\IP{\rlx\hbox{\rm I\kern-.18em P}}
\def\IQ{\rlx\hbox{\,$\inbar\kern-.3em{\rm Q}$}}
\def\IF{\rlx\hbox{\rm I\kern-.18em F}}
\def\IG{\rlx\hbox{\,$\inbar\kern-.3em{\rm G}$}}
\def\IH{\rlx\hbox{\rm I\kern-.18em H}}
\def\II{\rlx\hbox{\rm I\kern-.18em I}}
\def\IK{\rlx\hbox{\rm I\kern-.18em K}}
\def\IL{\rlx\hbox{\rm I\kern-.18em L}}
\def\one{\hbox{{1}\kern-.25em\hbox{l}}}
\def\0#1{\relax\ifmmode\mathaccent"7017{#1}%
B        \else\accent23#1\relax\fi}

\def\EPJC#1#2#3{{\sl Eur. Phys. J.} {\bf C#1} (#2) #3}
\def\EPJA#1#2#3{{\sl Eur. Phys. J.} {\bf A#1} (#2) #3}

                \def\JHEP#1#2#3{{\sl JHEP} {\bf#1} (#2) #3}
                \def\PRL#1#2#3{{\sl Phys. Rev. Lett.} {\bf#1} (#2) #3}
                \def\NPB#1#2#3{{\sl Nucl. Phys.} {\bf B#1} (#2) #3}

                \def\PRD#1#2#3{{\sl Phys. Rev.} {\bf D#1} (#2) #3}

                \def\PLB#1#2#3{{\sl Phys. Lett.} {\bf #1B} (#2) #3}
                \def\JMP#1#2#3{{\sl J. Math. Phys.} {\bf #1} (#2) #3}

                \def\AoP#1#2#3{{\sl Annals Phys. NY} {\bf #1} (#2) #3}

                \def\PR#1#2#3{{\sl Phys. Reports} {\bf #1} (#2) #3}

                \def\IJMPA#1#2#3{{\sl Int. J. Mod. Phys.} {\bf A#1}
(#2) #3}

\def\PAN#1#2#3{{\sl Phys.\ Atom.\ Nucl.} {\bf #1} (#2) #3}
\def\AIP#1#2#3{{\sl AIP Conf. Proc.} {\bf #1} (#2) #3}
\def\ZPA#1#2#3{{\sl Z.Phys.} {\bf A#1} (#2) #3}
\def\PD#1#2#3{{\sl Physica} {\bf D#1} (#2) #3}

                \def\a{\alpha}
                \def\b{\beta}

                \def\d{\delta}
                \def\D{\Delta}
                
                \def\g{\gamma}
                
                \def\vp{\varphi}

                \def\/{\frac}

                \def\l{\lambda}
                \def\L{\Lambda}
                
                \def\m{\mu}

                \def\pa{\partial}

                \def\vp{\varphi}
                
                \def\s{\sigma}

                \def\({\Big(}
                \def\){\Big)}
                \def\[{\Big[}
                \def\]{\Big]}

  \def\rlx{\relax\leavevmode}
                \def\inbar{\vrule height1.5ex width.4pt depth0pt}
                \def\IZ{\rlx\hbox{\sf Z\kern-.4em Z}}
                \def\IR{\rlx\hbox{\rm I\kern-.18em R}}
                \def\IC{\rlx\hbox{\,$\inbar\kern-.3em{\rm C}$}}
                \def\IN{\rlx\hbox{\rm I\kern-.18em N}}
                \def\IO{\rlx\hbox{\,$\inbar\kern-.3em{\rm O}$}}
                \def\IP{\rlx\hbox{\rm I\kern-.18em P}}
                \def\IQ{\rlx\hbox{\,$\inbar\kern-.3em{\rm Q}$}}
                \def\IF{\rlx\hbox{\rm I\kern-.18em F}}
                \def\IG{\rlx\hbox{\,$\inbar\kern-.3em{\rm G}$}}
                \def\IH{\rlx\hbox{\rm I\kern-.18em H}}
                \def\II{\rlx\hbox{\rm I\kern-.18em I}}
                \def\IK{\rlx\hbox{\rm I\kern-.18em K}}
                \def\IL{\rlx\hbox{\rm I\kern-.18em L}}
                \def\one{\hbox{{1}\kern-.25em\hbox{l}}}
                \def\0#1{\relax\ifmmode\mathaccent"7017{#1}%
                B        \else\accent23#1\relax\fi}
                
\vspace{.2in}
\begin{document}
\vspace{.2in}
\begin{center}
{\large\bf Exotic baryons in two-dimensional QCD and the
generalized sine-Gordon solitons}
\end{center}

\vspace{1in}


\begin{center}

Harold Blas  

\vspace{.5 cm} \small

\par \vskip .1in \noindent

               Departamento de Matem\'atica e F\'{\i}sica - ICET\\
Universidade Federal de Mato Grosso\\
 Av. Fernando Correa, s/n, Coxip\'o \\
78060-900, Cuiab\'a - MT - Brazil\\

\normalsize
\end{center}
\vspace{1 cm}

\begin{abstract}
The solitons and kinks of the $SU(3)$ generalized sine-Gordon
model (GSG) are  shown to describe the baryonic spectrum of
two-dimensional quantum chromodynamics (QCD$_{2}$). The GSG model
arises in the low-energy effective action of bosonized QCD$_{2}$
with unequal quark mass parameters. The GSG potential for
$N_{f}=3$ flavors resembles the potential of the effective chiral
lagrangian proposed by Witten to describe low-energy behavior of
four dimensional QCD. Among the attractive features of the GSG
model are the variety of soliton and kink type solutions for
QCD$_{2}$ unequal quark mass parameters [JHEP(0701)(2007)(027)].
Exotic baryons in QCD$_{2}$ [J. Ellis et al JHEP0508(2005)081] are
discussed in the context of the GSG model. Various semi-classical
computations are performed improving the results of this reference
and clarifying the role of unequal quark masses. The remarkable
double sine Gordon model also arises as a reduced GSG model
bearing a kink($K$) type solution describing a multi-baryon; so,
the description of some resonances in QCD$_{2}$ may take advantage
of the properties of the $K\bar{K}$ system.
\end{abstract}
\newpage





\par \vskip .3in \noindent
\vspace{1 cm}
\section{Introduction}
Quantum Chromodynamics in two-dimensions (QCD$_{2}$) (see e.g.
\cite{abdalla}) has long been considered a useful theoretical
laboratory for understanding
 non-perturbative strong-interaction problems such as confinement
 \cite{gross}, the large-N$_{c}$ expansion \cite{'tHooft},  baryon structure
\cite{frishman} and, more recently, the chiral-soliton picture for
normal and exotic baryons \cite{ellis}. Even though there are
various differences between QCD$_{4}$ and QCD$_{2}$, this theory
may provide interesting insights into the physical
four-dimensional world. In two dimensions, an exact and complete
bosonic description exists and in the strong-coupling limit one
can eliminate the color degrees of freedom entirely, thus getting
an effective action expressed in terms of flavor degrees of
freedom only. In this way various aspects have been studied, such
as baryon spectrum and its $\bar{q}q$ content \cite{frishman}. The
constituent quark solitons of baryons were uncovered taking into
account the both bosonized flavor and color degrees of freedom
\cite{ellis1}. In particular, the study of meson-baryon scattering
and resonances is a nontrivial task for unequal quark masses even
in 2D \cite{ellis2}.

Recently, in QCD$_{4}$ there appeared some puzzles related with
unequal quark masses \cite{exp}
 providing an extra motivation to consider
QCD$_{2}$ as a testing ground for non-perturbative methods that
might have relevance in the real world. Claims for the existence
of exotic baryons - that can not be composed of just three quarks
- have inspired intense studies of the theory and phenomenology of
QCD in the strong-interaction regime. In particular, it has led to
the discovery that the strong coupling regime may contain
unexpected correlations among groups of two or three quarks and
antiquarks. Results of growing number of experiments at
laboratories around the world provide contradictive
 situation regarding the experimental observation of possible pentaquark states, see
e.g. \cite{schumacher}. These experiments have thus opened new
lines of theoretical investigation that may survive even if the
original inspiration - the exotic $\Theta^{+}$ pentaquark
existence- is not confirmed. After the reports of null results
started to accumulate the initial optimism declined, and the
experimental situation remains ambiguous to the present. The
increase in statistics led to some recent new claims for positive
evidence \cite{barmin}, while the null result \cite{mckinnon} by
CLAS is specially significant because it contradicts their earlier
positive result, suggesting that at least in their case the
original claim was an artifact due to low statistics. All this
experimental activity spurred a great amount of theoretical work
in all kinds of models for hadrons and a renewed interest in
soliton models. Recently, there is new strong evidence of an
extremely narrow $\Theta^{+}$ resonance from DIANA collaboration
and a very significant new evidence from LEPS. According to
Diakonov, ``the null results from the new round of CLAS
experiments are compatible with what one should expect based on
the estimates of production cross sections" \cite{diakonov1}.

It has been conjectured that the low-energy action of QCD$_{2}$
($e_{c}>>M_{q}$, $M_{q}$ quark mass and $e_{c}$ gauge coupling)
 might be related to massive two dimensional integrable models, thus leading to the exact solution
 of the strong coupled  QCD$_{2}$ \cite{frishman}. As an example of this
picture, it has been shown that the so-called  $su(2)$ affine Toda
model coupled to matter (Dirac) field (ATM) \cite{matter}
describes the low-energy spectrum of QCD$_{2}$ (one flavor and
$N_{C}$ colors) \cite{prd}. The ATM model allowed the exact
computation of the string tension in QCD$_{2}$ \cite{prd},
improving the approximate result of \cite{armoni}. The strong
coupling sector of the $su(2)$ ATM model is described by the usual
sine-Gordon model \cite{nucl1, nucl, annals}. The baryons in QCD
may be described as solitons in the bosonized formulation. In the
strong-coupling limit the static classical soliton which describes
a baryon in QCD$_{2}$ turns out to be the ordinary sine-Gordon
kink, i.e. \br \Phi(x) = \frac{4}{\b_{0}}
\mbox{tan}^{-1}[\mbox{exp}\, \b_{0} \sqrt{2} \widetilde{m} x]
\label{soliton}\er where $\b_{0} = \sqrt{\frac{4\pi}{N_{C}}}$ is
the coupling constant of the sine-Gordon theory,
$8\sqrt{2}\widetilde{m}/\b_{0}$  is the mass of the soliton, and
$\widetilde{m}$ is related to the common bare mass of the quarks
by a renormalization group relation relevant to two dimensions.
The soliton in (\ref{soliton}) has non-zero baryon number as well
as Y charge. The quantum correction to the soliton mass, obtained
by time-dependent rotation in flavor space, is suppressed by a
factor of $N_{C}$ compared to the classical contribution to the
baryon mass \cite{frishman}. The considerations of more
complicated mass matrices and higher order corrections to the
$M_{q}/e_{c} \rightarrow 0$ limit are among the issues that
deserve further attention.

In this context, we show that various aspects of the low-energy
effective QCD$_{2}$ action with unequal quark masses can be
described by the (generalized) sine-Gordon model (GSG). The GSG
model has appeared in the study of the strong coupling sector of
the $sl(n,\IC)$ ATM theory\cite{jmp, jhep, jhep4}, and in the
bosonized multiflavor massive Thirring model \cite{epjc}.  In
particular, the GSG model
 provides the framework to obtain (multi-)soliton solutions for unequal quark mass
parameters. Choosing the normalization such that quarks have
baryon number $Q_{B}^{0}=1$ and a one-soliton has baryon number
$N_{C}$, we classify the configurations in the GSG model with
baryon numbers $N_{C},\,2N_{C},\,...\, 4N_{C}$. For example, the
double sine-Gordon model provides a kink type solution describing
a multi-baryon state with baryon number $4N_{C}$ (see Appendix).
Then, using the GSG model we generalize the results of refs.
\cite{frishman, ellis} which applied the semi-classical
quantization method in order to uncover the normal \cite{frishman}
and exotic baryon \cite{ellis} spectrum of QCD$_{2}$. One of the
main features of the GSG model is that the one-soliton solution
requires the QCD quark mass parameters to satisfy certain
relationship. In two dimensions there are no spin degrees of
freedom, so, the lowest-lying baryons are related to the purely
symmetric Young tableau, the ${\bf 10}$ dimensional representation
of flavor $SU(3)$. This is the analogue of the multiplet
containing the baryons $\Delta,\, \Sigma,\, \Xi,\, \Omega^{-}$ in
QCD$_{4}$. The next state corresponds to a state with the quantum
numbers of four quarks and an antiquark, the so-called pentaquark,
which in two dimensions forms a ${\bf 35}$ representation of
flavor $SU(3)$. This corresponds to the four dimensional multiplet
$\overline{\bf 10}$, which contain the exotic baryons
$\Theta^{+},\,\bar{\Sigma},\, \Xi^{--}$.

Here we improve the results of refs. \cite{frishman, ellis}, such
as the normal and exotic baryon masses, the relevant mass ratios
and the radius parameter of the exotic baryons. The semi-classical
computations of the masses get quantum corrections due to the
unequal mass term contributions and to the form of the diagonal
ansatz taken for the flavor field (related to GSG model)
describing the lowest-energy state of the effective action. The
corrections to the normal baryon masses are an increase of $3.5\%$
to the earlier value obtained in \cite{ellis}, and in the case of
the exotic baryon our computations improve the behavior of the
quantum correction by decreasing the earlier value in $0.34$
units, so making the semi-classical result more reliable. Let us
mention that for the first exotic baryon \cite{ellis} the quantum
correction was greater than the classical term by a factor of
$2.46$, so that semi-classical approximation may not be a good
approximation. As a curiosity, with the relevant values obtained
by us for QCD$_{2}$ we computed the ratio between the lowest
exotic baryon and the $R={\bf 10}$ baryon masses \, $M_{{\bf
35}}/M_{{\bf 10}} \sim 1.65$, which is only $1\%$ larger than the
analogous  four dimensional QCD ratio $M_{\Theta^{+}}/M_{nucleon}
\sim 1.63$. In \cite{ellis} the relevant QCD$_{2}$ ratio was
$17\%$ larger than this value. The mass formulae for the normal
and exotic baryons corresponding, respectively, to the
representations ${\bf 10}$ and ${\bf 35}$, in two dimensions
resemble the general chiral-soliton model formula in four
dimensions \cite{Diakonov1} except that there is no spin-dependent
term $\sim  J(J+1)$, and an analog term containing the soliton
moment of inertia emerges.

The paper is organized as follows. The next section summarizes the
bosonized low-energy effective action of QCD$_{2}$ and introduces
the lowest-energy state described by the GSG action. The global
 QCD$_{2}$ symmetries are discussed. In section \ref{semicl} the
 semi-classical method of quantization relevant to a general diagonal ansatz is introduced. In
subsection  \ref{review} we briefly review the ordinary
sine-Gordon soliton semi-classical quantization in the context of
QCD$_{2}$. In section \ref{gsgqcd} we discuss the quantum
correction to the $SU(3)$ GSG ansatz in the framework of
semi-classical quantization. In subsection \ref{nc} the GSG
one-soliton state is rotated in $SU(3)$ flavor space by a time-
dependent $A(t)$. In subsection \ref{lowest} the lowest-energy
baryon state with baryon number $N_{C}$ is introduced. The
possible vibrational modes are briefly discussed in subsection
\ref{vibr}. Section \ref{exo} discusses the first and higher
multiplet exotic baryons and provides the relevant quantum
corrections the their masses, the ratio  $M_{{\bf 35}}/M_{{\bf
10}}$, and an estimate for the exotic baryon radius parameter. The
last section presents a summary and some discussions. In the
appendix we provide the GSG solitons and kink solutions relevant
to our
 QCD$_{2}$ discussion.

\section{Baryons in Bosonized QCD$_{2}$}
\subsection{The bosonized effective action}

The QCD$_2$ action is written in terms of gauge fields $A_\mu$ and fundamental quark
fields $\psi$ as
\begin{equation}
S_F[\psi,A_\mu]=\int d^2x\{-\frac{1}{2 e_c^2} Tr(F_{\mu\nu} F^{\mu\nu})-\bar
\psi^{ai} {[(i\sp +\sa) ]\psi_{a i}} + {\cal M}_{ij} \bar \psi^{ai} \psi_{aj} \} ,
\label{action}
\end{equation}
where $a$ is the color index ($a=1,2,...,N_{C}$) and $i$ the
flavor index ($i=1,2,..,N_{f}$), $e_c$, with dimension of a mass,
is the quark coupling to the gauge fields, the matrix ${\cal
M}_{ij}=m_{i} \d_{ij}$ ($m_{i}$ being the quark masses) takes into
account  the quark  mass splitting, and $F_{\mu\nu} \equiv
\partial_\mu A_\nu -\partial_\nu A_\mu +i[A_\mu ,A_\nu ] $ is the gauge field
strength.

The bosonized action in the strong-coupling limit ($e_{c} >>$ all $m_{i}$)  becomes
\cite{frishman, ellis1}
\begin{equation}
S_{eff}[g] =N_c S[g]+m^2N_m\int d^2x Tr_{f}\[ {\cal D} (g + g^{\dagger})\],
\label{Baction}
\end{equation}
where $g$ is a matrix representing $U(N_f)$, ${\cal D}= \frac{\cal M}{m_{0}}$,
\,$m_{0}$ is an arbitrary mass parameter and the effective mass scale $m$ is given by
\begin{equation}
m =[N_c c m_{0} ({e_c\sqrt{N_F}\over \sqrt{2\pi}})^{\Delta_{c}} ]^{1\over
1+\Delta_{c}}, \label{mass}
\end{equation}
with
\begin{equation}
\Delta_c = {N_c^2-1\over N_c(N_c+N_F)}. \label{Deltac}
\end{equation}

In (\ref{Baction}) $S[g]$ is the WZNW action and $N_{m}$ stands
for normal ordering with respect to $m$. In the large $N_c$ limit,
which we use below to justify the semi-classical approximation,
the scale $m$ becomes \br\label{mlargen} m \,=\, 0.59 N_F ^{1\over
4}\, \sqrt {N_c e_c m_{0}},\er so, $m$ takes the value\,
$0.77\sqrt {N_c e_c m_{0}}$\, for three flavors. Notice that we
first take the strong-coupling limit $e_c \gg $\, all \, $m_i$,
and then take $N_c$ to be large, thus it is different from the 't
Hooft limit \cite{'tHooft}, where $e_c^2 N_c$ is held fixed.

Following the  Skyrme model approach it is useful to first ask for
classical soliton solutions of the bosonic action which are heavy
in the $N_{C}\rightarrow $large limit. The action (\ref{Baction})
is a massive WZNW action and possesses
 the property that if $g$ is non-diagonal it can not be a
classical solution, as after a diagonalization to \br
\label{diag0} g_{0}= \mbox{diag}(e^{-i \b_{0} \Phi_{1}(x)},\,
e^{-i\b_{0}
\Phi_{2}(x)},\,...\,e^{-i\b_{0}\Phi_{N_{f}}(x)}),\,\,\,\,\,\,
\sum_{i} \Phi_{i}(x)=\phi(x), \,\,\,\,\,\,\,\b_{0}\equiv
\sqrt{\frac{4\pi}{N_{C}}}\er it will have lower energy
\cite{Frishman:1997}. Thus, the minimal energy solutions of the
massive WZNW model are necessarily in a diagonal form. The
majority of particles given  by (\ref{diag0}) are not going to be
stable, but must decay into others.

Previous works consider the diagonal form (\ref{diag0}) such that
the action (\ref{Baction}) reduces to a sum of $N_{f}$ {\sl
independent} ordinary sine-Gordon models, each one for the
corresponding $\Phi_{i}$ field and parameters \br
\widetilde{m}_{i}^2 = \frac{m_{i}}{m_{0}} m^2.\label{masspar}\er

In this approach the lowest lying baryon is represented by the
minimum-energy configuration for this class of ansatz, i.e. \br
\label{diag1} \hat{g}_{0}(x)= \mbox{diag}\(1,\,1,\,...,\,e^{-i
\sqrt{\frac{4\pi}{N_{C}}} \Phi_{N_{f}}}\),\er with $m_{N_{f}}$
chosen to be the smallest mass.

In this paper we will consider the ansatz (\ref{diag0}) for  \br
N_{f} = \frac{n}{2} (n-1),\,\,\,\,\, N_{f}\equiv \mbox{number of
positive roots of} \,\,su(n),\er such that $\frac{(n-2)(n-1)}{2}$
linear constraints are imposed on the fields $\Phi_{i}$. This
model corresponds to the generalized sine-Gordon model (GSG)
recently studied in the context of the bosonization of the
so-called generalized massive Thirring model (GMT) with $N_{f}$
fermion species \cite{jmp, jhep, epjc}. The classical GSG model
and some of its properties, such as the algebraic construction
based on the affine $sl(n, \IC)$ Kac-Moody algebra and the soliton
spectrum has been the subject of a recent paper \cite{jhep4}.

The WZ term in (\ref{Baction}) vanishes for either static or
diagonal solution, so, for the ansatz (\ref{diag0}) and after
redefining the additive constant term the action becomes \br
\label{GSG} S[g_{0}]= \int d^2x \sum_{i=1}^{N_{f}}\[ \frac{1}{2}
(\pa_{\mu} \Phi_{i})^2 + 2 \widetilde{m}_{i}^2 \( \mbox{cos}
\b_{0} \Phi_{i}-1\)\], \er with coupling $\b_{0}$ and mass
parameters $\widetilde{m}_{i}$ defined in (\ref{diag0}) and
(\ref{masspar}), respectively.

The $\Phi_{i}$ fields in (\ref{GSG}) satisfy certain constraints
of the type \br \label{const} \Phi_{p} = \sum_{i=1}^{n-1}\s_{pi}
\Phi_{i},\,\,\,\,\,p=n, n+1, ..., N_{f}\er where $\s_{p\,i}$ are
some constant parameters. From the Lie algebraic construction of
the GSG model these parameters arise from the relationship between
the positive and simple roots of $su(n)$. Even though our
treatment until section \ref{semicl} is valid for any $N_{f}$,
starting in section \ref{gsgqcd} we will concentrate on the
$N_{f}=3$ case.

It is interesting to recognize the similarity between the
potential of the model (\ref{GSG})-(\ref{const}) for the $N_{f}=3$
case [in su(3) GSG model one has $n = N_{f} = 3$ and just one
 constraint equation in (\ref{const})] and the effective chiral
Lagrangian proposed by Witten to describe low-energy behavior of
four dimensional QCD \cite{witten}. In Witten's approach  the
potential term reads \br\label{witten} V^{\mbox{Witten}}(U) =
f_{\pi}^2 \Big[ -\frac{1}{2} Tr M (U+U^{\dagger}) + \frac{k}{
2N_{C}} (-i\mbox{ln} \mbox{Det}\, U -\theta )^2\Big], \er where
$U$ is the pseudoscalar field matrix and $M = \mbox{diag}
\(m_{u};m_{d};m_{s}\)$ is the quark mass matrix.
Phenomenologically $ m^2_{\eta'} >>
m^2_{\pi},\,m^2_{K},\,m^2_{\eta}$, implying that $\frac{k}{N_{C}}
> b\, m_{s}>> b\, m_{u},\,b\, m_{d}\,$ [the parameter $b$ is $O(\L)$, where
$\L$ is a hadronic scale). Because $M$ is diagonal, one can look for a
minimum of $V^{\mbox{Witten}}(U)$
 in the form $U = \mbox{diag} \(e^{i\phi_{1}},\,
e^{i\phi_{2}},\, e^{i\phi_{3}}\)$. Since the second term dominates
over the first, one has $\sum \phi_{j} = \theta$ up to the first
approximation. So, choosing $\theta=0$, (\ref{witten}) reduces to
a model of type  (\ref{GSG})-(\ref{const}) defined for $N_{f}=3$.
This is the $sl(3)$ GSG model, which possesses soliton and kink
type solutions (see the Appendix), and will be the main ingredient
of our developments in sections \ref{gsgqcd} and \ref{exo}.

The potential term in (\ref{GSG}) is invariant under
                \br \label{vac} \Phi_{i} \rightarrow \Phi_{i} + \frac{1}{\b_{0}} 2\pi N_{i},\,\,(N_{i} \in
                \IZ).\er

All finite energy configurations, whether static or
time-dependent, can be divided into an infinite number of
topological sectors, each characterized by a set \br \[n_{1},
n_{2}, ..., n_{N_{f}}\] &=& \label{indices}
                \[(N_{1}^{+}-N_{1}^{-}), (N_{2}^{+}-N_{2}^{-}), ..., (N_{N_{f}}^{+}-N_{N_{f}}^{-})\]\\
                \Phi_{i}(\pm \infty)&=& \frac{1}{\b_{0}} 2\pi N_{i}^{\pm} \er
corresponding to the asymptotic values of the fields at $x= \pm
\infty$. The $n_{i}'s$ satisfy certain relationship arising from
the constraints (\ref{const}) and the invariance (\ref{vac}) (some
examples are given in the appendix for the soliton and kink  type
solutions in the $SU(3)$ case).

Conserved charges, corresponding to the vector current $
J_{ij}^{\mu}= \bar{\psi}^{a}_{i}\g^{\mu}\psi_{j}^{a}$, can be
computed as \br Q^{A}[g(x)]= \int dx [J_{0} (\frac{T^{A}}{2})],
\label{charge0}\er where $(\frac{T^{A}}{2})$ are the $su(n)$
generators and the $U(1)$ baryon number is obtained using the
identity matrix instead of $(\frac{T^{A}}{2})$. For $g_{0}$ given
in  eq. (\ref{diag0}) the baryon number of any given flavor $j$ is
given by ${\cal Q}_{B}^{j}=N_{c} n_{j}$, so, the total baryon
number
 becomes \br \label{baryon} {\cal Q}_{B}&=& N_{C} (n_{1}+
n_{2}+...+n_{N_{f}}),\er and the  ``hypercharge" is given by \br
\nonumber Q_{Y}&=& \frac{1}{2} \mbox{Tr} \int dx \(J_{0}
\l_{N_{f}^2 -1} \)
\\\label{hyper} &=&\frac{1}{2} N_{C} \(n_{1}+
n_{2}+...+n_{N_{f}-1} - (N_{f}-1) n_{N_{f}}\)
\sqrt{\frac{2}{N_{f}^2 - N_{f}}}.\er

The total baryon number is clearly an integer multiple of $N_{C}$.
In the case of (\ref{diag1}) they reduce to ${\cal Q}_{B}=N_{C}$
and $Q_{Y}^{0}=- \frac{1}{2} \sqrt{2(N_{f}-1)/N_{F}} N_{C}$,
respectively [for $\sqrt{4\pi/N_{C}}
\Phi_{N_{f}}(+\infty)=2\pi,\,\,\Phi_{N_{f}}(-\infty)=0$]
\cite{frishman}. We are choosing the convention in which
 the quarks have baryon number $Q_{B}=1$, so the soliton representing  a physical baryon has
 baryon number $N_{C}$.

A global $U_{V}(N_{f})$ transformation $\widetilde{g}_{0}= A g_{0}(x) A^{-1}$ is
expected to turn on the other charges. Let us introduce
 \br\label{matrix} A &=& \Bigg( \begin{array}{ccc}
 z_{1}^{(1)}& ...& z_{1}^{(N_{f})} \\ z_{2}^{(1)}& ...& z_{2}^{(N_{f})}
\\z_{N_{f}}^{(1)}& ...& z_{N_{f}}^{(N_{f})}\end{array}\Bigg),\\
&& \sum_{p =1}^{N_{f}} z_{p}^{(i)}z_{p}^{(j)\, \star}=\d_{ij}.
\label{compl2}
 \er

Now
\br \widetilde{g}_{0}= \sum_{j=1}^{N_{f}} e^{i \b_{j}
\Phi_{j}} \IZ^{(j)},\,\,\,\,\,\,\,\,\,  \IZ^{(j)}_{p\,q} =
z_{p}^{(j)}z_{q}^{(j)\, \star},\label{expan}\er

The charges with $\widetilde{g}_{0}$ are \br (\widetilde{Q}^{0})^{A}=\frac{1}{2}
N_{C} \mbox{Tr} \sum_{i} \( n_{i} T^{A}\IZ^{(i)}\)
 \er

 The baryon number is unchanged.  The $U(n)$ possible representations will be
 discussed below in the semi-classical quantization approach.

 \section{Semi-classical quantization and the GSG ansatz}
\label{semicl}

In order to implement the semi-classical quantization let us
consider \br \label{rota1} g(x,t )= A(t) g_{0}(x)
A^{-1}(t),\,\,\,\,\,\,A(t) \in U(N_{f})\er and derive the
effective action for $A(t)$ by substituting $g(x,t)$ into the
original action. So, following similar steps to the ones developed
in \cite{frishman} one can get

\br \widetilde{S}(g(x,t))-\widetilde{S}(g_{0}(x))= \frac{N_{C}}{8\pi}\int d^2x
\mbox{Tr}\(\[A^{-1}\dot{A}\,,\,g_{0}\]\[A^{-1}\dot{A}\,,\,g_{0}^{\dagger}\]\)+\nonumber \\
\frac{N_{C}}{2\pi}\int d^2x \mbox{Tr}\{(A^{-1}\dot{A}) (g_{0}^{\dagger}
\pa_{x}g_{0})\}+ m^2 \mbox{Tr} \int d^2x[({\cal D} Ag_{0}A^{\dagger} - {\cal D} g_{0})+ c.c. ]\label{g0}\er

The action above for ${\cal D}_{ij} = \d_{ij}$ (in this case the
last integrand after taking the trace operation vanishes
identically) is invariant under global $U(N_{f})$ transformation
\br A \rightarrow U A,\label{symm00}\er where $U \in G= U(N_{f})$.
This corresponds to the invariance of the original action (with
mass of the same magnitude for all flavors) under $g \rightarrow
UgU^{-1}$. It is also invariant under the local changes \br A(t)
\rightarrow A(t) V(t)\label{symm11},\er where $V(t) \in H $. This
subgroup $H$ of $G$ is nothing but the invariance group of
$g_{0}$. Below we will find some particular cases of $H$.

We define the Lie algebra valued variables $q^{i},\, y_{a}$
through $A^{-1} \dot{A} = i \sum \{\dot{q}_{i} E_{\a_{i}} +
\dot{y}_{a} H^{a}\}$ in the generalized Gell-Mann representation
\cite{bertini}. In terms of these variables the action (\ref{g0}),
for a diagonal mass matrix such that ${\cal D}_{ij} = \d_{ij}$,
takes the form \br S[q, y]= \int dt \{ \sum_{i=1}^{N_{f}}
\frac{1}{2M_{i}}\dot{q}_{i}\, \dot{q}_{-i}
- \sum_{a=1}^{N_{f}-1} \sqrt{\frac{2}{(a+1)^2- (a+ 1)}} \, \times \nonumber \\
\times\, (n_{1}+ n_{2}+...+n_{a}- a n_{a+ 1})\, \dot{y}_{a}\}
\label{g01},\er where $q_{\pm i}$ are associated to the positive
and negative roots, respectively, and \br \frac{1}{2M_{i}} =
\frac{N_{C}}{2\pi} \int_{-\infty}^{+\infty} [1- \mbox{cos} \b_{0}
\Phi_{i}],\,\,\,\,\,  \Phi_{i} \neq 0. \label{mm1}\er

In the case of vanishing $\Phi_{j} \equiv 0$ for a given $j$ one
must formally set $M_{j}=+ \infty$ in the relevant terms
throughout.


In the case of $\hat{g}_{0} = \mbox{diag}(1,1,...,
e^{i\b_{N_{f}}\Phi_{N_{f}}})$ the second summation  in (\ref{g01})
reduces to the unique term $[- N_{c}
\sqrt{\frac{2(N_{f}-1)}{N_{f}}}\, \dot{y}_{N_{f}-1} ]$
\cite{frishman}.

When written in terms of the general diagonal field $g_{0}(x)$ and
 the $U(N_{f})$ field $A(t)$, the charges associated to the global $U(N_{f})$ symmetry,
(\ref{charge0}), become \br \label{charge11} Q^{B}=
i\frac{N_{C}}{8\pi} \int dx \mbox{Tr} \{T^{B} A\{(g_{0}^{\dagger}
\pa_{x} g_{0}-g_{0} \pa_{x} g_{0}^{\dagger})+ [g_{0}\,,\,
[A^{-1}\dot{A}\,,\, g_{0}^{\dagger}]]\}A^{-1}\} \er

A convenient parameterization, instead of the parameters used in
(\ref{g01}), is (\ref{matrix}) since in the above expressions, for
$Q^{B}$ and the action (\ref{g0}), there appear the fields $A,
A^{-1}$, as well as
 their time derivatives. Now, for a diagonal mass matrix such
that ${\cal D}=\frac{m_{i}}{m_{0}} \d_{ij}$, the expression
(\ref{g0}) can be written  in terms of the variables
 $z_{p}^{(i)}$ subject to the relationships (\ref{compl2})
 \br
 \widetilde{S}(g(x,t))-\widetilde{S}(g_{0}(x))= S[z_{p}^{(i)}(t), \Phi_{i}(x)]
\er
\br
 S[z_{p}^{(i)}(t), \Phi_{i}(x)]&=& \frac{N_{C}}{2\pi} \int d^2 x \sum_{p, q;\, i<j}
 [\mbox{cos}(\b_{i}\Phi_{i}-\b_{j}\Phi_{j})-1] [ \dot{z}^{(i)}_{p} z^{(i)\,\star}_{q}
 \dot{z}^{(j)}_{q} z^{(j)\,\star}_{p}] -\nonumber\\&&
i\frac{N_{C}}{2\pi} \int d^2 x \sum_{i, p}\b_{i} \pa_{x} \Phi_{i}
\dot{z}_{p}^{(i)}z_{p}^{(i)\, \star} +\nonumber\\ && \int dt\, 2
[\sum_{i, p} \mbox{cos} (\b_{i}\Phi_{i}) \widetilde{m}^2_{p}
z_{p}^{(i)} z_{p}^{(i)\, \star} - \sum_{i} \mbox{cos}
(\b_{i}\Phi_{i}) \widetilde{m}_{i}^2] \label{fieldvp} \er

Let us choose the index $k$ corresponding to the smallest mass
$m_{k}$. So, integrating over $x$ in (\ref{fieldvp}) we may write
\br\nonumber
 S[z_{p}^{(i)}(t)]&=&
  -\frac{1}{2} \int dt \, \sum_{i<j}^{N_{f}} \sum_{p, q}^{N_{f}} M_{ij}^{-1} \dot{z}^{(i)}_{p} z^{(i)\,\star}_{q}
  \dot{z}^{(j)}_{q} z^{(j)\,\star}_{p} - \\
&& i \frac{N_{C}}{2} \int dt  \sum_{i} n_{i} \[ \dot{z}_{p}^{(i)}
z^{(i)\,\star}_{p}-  z_{p}^{(i)} \dot{z}^{(i)\,\star}_{p}\]-
\nonumber\\&& \frac{2\pi}{N_{c}} \int dt \Big\{ \sum_{i,\,p}
\[ \frac{\widetilde{m}^2_{p}}{M_{i}} - \frac{\widetilde{m}^2_{k}}{M_{k}}\] z_{p}^{(i)}
z_{p}^{(i)\,\star} + \frac{2\pi}{N_{c}}
\[ \sum_{i} \frac{\widetilde{m}^2_{i}}{M_{i}} - \frac{\widetilde{m}^2_{k}}{M_{k}} N_{\phi} \] \Big\}\nonumber
\\ &&
 +\int dt  (z_{p}^{(i)} z^{(j)\,\star}_{p}-\d_{ij}) \l_{ij} \label{lag1} \er where $N_{\phi}$ is the number of
 nonvanishing $\Phi_{i}$ fields and we have introduced some Lagrange multipliers
enforcing the relationships (\ref{compl2}). The constants $M_{ij}$
above are defined by \br \frac{1}{2 M_{ij}} &\equiv&
\frac{N_{C}}{2\pi} \int dx
[1-\mbox{cos}(\b_{0}\Phi_{i}-\b_{0}\Phi_{j})]; \,\,\,\,\,\,\,\, i<
j. \label{mij}\er

If the field solutions are such that $\Phi_{i}=\Phi_{j}$, then one
must set formally $M_{ij} \rightarrow +\infty$ in place of the
corresponding constants.

Likewise, we can write the $U(N_{f})$ charges, eq.
(\ref{charge11}), in terms of the $z_{p}^{(i)}$ variables
\br Q^{A}&=& \frac{1}{2} T^{A}_{\b \a} Q_{\a\b},\nonumber\\
\label{charge22} Q_{\a\b} &=& N_{C} \sum_{j} n_{j}\, z_{\a}^{(j)}
z_{\b}^{(j)\, \star} - \frac{i}{2} \sum_{i,j}  M^{-1}_{ij}
z_{\a}^{(j)} z_{\g}^{(j)\, \star} \dot{z}_{\g}^{(i)}
z_{\b}^{(i)\,\star}.\er

The second $U(N_{f})$ Casimir operator is obtained from the charge
matrix elements $Q_{\a\b}$

\br Q^{A} Q_{A}&=& \frac{1}{2} Q_{\a\b}Q_{\b \a},\nonumber\\
&=& \frac{1}{2}N_{C}^2 \sum_{i} n_{i} n_{i} - \frac{1}{4}
\sum_{i<j} \(M_{ij}^{-1} \)^2 \dot{z}_{\a}^{(j)} z_{\b}^{(j)\,
\star} \dot{z}_{\b}^{(i)} z_{\a}^{(i)\,\star}\label{casimir1}\er

The expressions above greatly simplify in certain particular cases
of the ansatz (\ref{diag0}), the ansatz (\ref{diag1}) has been
studied extensively in the literature before. In the next
subsection we review this case and in further sections we analyze
the semiclassical quantization of the GSG ansatz given for
$N_{f}=3$ flavors.

\subsection{Review of usual sine-Gordon soliton and baryons in
QCD$_{2}$} \label{review}

In this subsection we briefly review the formalism applied to the
ansatz (\ref{diag1}), which is related to the usual SG one-soliton
as the lowest baryon state. In order to calculate the quantum
correction it is allowed the sine-Gordon soliton to rotate in
$SU(N_{f})$ space by a time dependent matrix $A(t)$ as in
(\ref{rota1}). Let us consider the single baryon state defined for
the ansatz (\ref{diag1}) for the sine-Gordon soliton solution
$\Phi_{N_{f}}\equiv \Phi_{1-soliton} $ [ $\Phi_{1-soliton}$ is
given by (\ref{soliton})]; so,  in the relations above one must
set \br n_{N_{f}}=1;\,\,\,\,n_{j}=0 \, (j\neq N_{f}); \,\,\,\,
M^{-1}_{j\,k}\equiv 0\,\, (j < k <
N_{f});\,\,\,\,M^{-1}_{j\,N_{f}} \equiv M^{-1}_{N_{f}}\,\, (j <
N_{f})\label{onebaryon},\er where $M^{-1}_{N_{f}}$ can be computed
using eq. (\ref{mm1}) for $i=N_{f}$ for the soliton
(\ref{soliton}) \br \label{mm0} \frac{1}{2M_{N_{f}}} =
\frac{1}{\sqrt{2}\, \widetilde{m}} (\frac{N_{C}}{\pi})^{3/2}\er

Then, for the ansatz ({\ref{diag1}}), i.e. $\hat{g}_{0}(x)=
\mbox{diag}\(1,\,1,\,...,\,e^{-i \sqrt{\frac{4\pi}{N_{C}}}
\Phi_{N_{f}}}\)$, the effective action (\ref{lag1}) can be written
as \br S[z^{(N_{f})}_{j}(t)] &=& \frac{1}{2M_{N_{f}}} \int \, dt
[\dot{z}_{j}^{(N_{f})\,\star}\dot{z}^{(N_{f})}_{j}-(z_{i}^{(N_{f})\,\star}
\dot{z}^{(N_{f})}_{i})(\dot{z}^{(N_{f})\,\star}_{k}
z^{(N_{f})}_{k})] \nonumber\\
\nonumber &&-\frac{2\pi}{M_{N_{f}} N_{C}} \int \, dt \,
\sum_{i=1}^{N_{f}} \(
\widetilde{m}^{2}_{i}-\widetilde{m}^{2}_{N_{f}}\) z_{i}^{(N_{f})\,
\star} z_{i}^{(N_{f})}
\\
&& -i \frac{N_{C}}{2} \int  dt\, n_{N_{f}}
\(z_{j}^{(N_{f})\,\star}\dot{z}^{(N_{f})}_{j}-\dot{z}_{j}^{(N_{f})\,\star}z^{(N_{f})}_{j}
\) \nonumber \\ && +\int dt [ ( z_{p}^{(N_{f})}
z^{(N_{f})\,\star}_{q}-\d_{pq}) \l^{pq} ] ,\label{lag2}\er where
$n_{N_{f}}=1$,\, $M_{N_{f}}$ is given by (\ref{mm0}) and
$m_{N_{f}}$ entering $\widetilde{m}_{N_{f}}$ is chosen to be the
smallest quark mass. Notice that for equal quark masses the second
line in eq. (\ref{lag2}) vanishes identically. According to
(\ref{symm00})-(\ref{symm11}), the symmetries of
$S[z^{(N_{f})}_{j}(t)]$  are the global $U(N_{f})$ group (for
equal quark masses) under which \br z^{(N_{f})}_{\a} \rightarrow
z^{'\,(N_{f})}_{\a}= U_{\a\b} z^{(N_{f})}_{\b} ,\,\,\,\,\,\,\,U
\in U(N_{f}),\er and a local $U(1)$ subgroup of $H$ under which
\br z^{(N_{f})}_{\a} \rightarrow z^{'\,(N_{f})}_{\a}= e^{i \d(t)}
z^{(N_{f})}_{\a}. \label{u1} \er

The action (\ref{lag2}) has been considered in order to find the
quantum correction to the soliton mass for certain representations
$R$ of the flavor symmetry $SU(N_{f})$. The case of equal quark
masses has  been studied in the literature \cite{frishman, ellis,
ellis2, frishman87}. Certain properties in the case of different
quark masses have been considered in \cite{ellis1} for the ansatz
(\ref{diag1}).

In this approach the minimum-energy configuration for the class of
ansatz (\ref{diag1}), with $m_{N_{f}}$ the smallest mass,
corresponds to the state of lowest-lying baryon \cite{frishman}
which in the large-$N_{C}$ limit possesses the classical mass
\br\label{baryonmass} M_{baryon}^{cl} = 4 \widetilde{m}_{N_{f}} \(
\frac{2N_{C}}{\pi}\)^{1/2} \approx 1.90 N_{f}^{1/4} \sqrt{e_{c}
m_{N_{f}}} N_{C},\er where $\widetilde{m}_{N_{f}}$ has been given
in (\ref{masspar}) for $i=N_{f}$.

Moreover, for the Ansatz (\ref{diag1}) the $SU(N_{f})$ charges
become \br Q_{\a\b} & = & N_{C} n_{N_{f}} z^{(N_{f})}_{\a}
z^{(N_{f})\,\star}_{\b} + \frac{i}{2M_{N_{f}}} \Big[
z^{(N_{f})}_{\a}z^{(N_{f})\,\star}_{\b}(\dot{z}^{(N_{f})}_{\d}
z^{(N_{f})\, \star}_{\d} - z^{(N_{f})}_{\d} \dot{z}^{(N_{f})\,
\star}_{\d})+ z^{(N_{f})}_{\a} \dot{z}^{(N_{f})\, \star}_{\b}\nonumber \\
&& -\dot{z}^{(N_{f})}_{\a} z^{(N_{f})\, \star}_{\b} \Big]
\label{charge33} \er

The corresponding second Casimir can be obtained from
(\ref{casimir1}) \br  Q_{A}Q^{A}= \frac{1}{2} Q_{\a \b} Q_{\b \a}
= \frac{1}{2} N_{C}^2 n^2_{N_{f}} + \frac{1}{4 M^{2}_{N_{f}}}
\(Dz\)^{\dagger}_{\a} \( D z\)_{\a},\,\,\,\,\,\,\,D z \equiv
\dot{z} - z (z^{\dagger} \dot{z})\label{casimir2}\er

Moreover, denoting the $SU(N_{f})$ second Casimir operator by
$C_{2}(N_{f})$ one can write \br Q_{A}Q^{A}= C_{2}(N_{f}) +
\frac{1}{2N_{f}} ({\cal Q}_{B})^2 \label{casimir00}, \er where
${\cal Q}_{B}$ is the baryon number (\ref{baryon}), which in this
case reduces to ${\cal Q}_{B}=N_{C}$.

In the case of a single baryon given by $\hat{g}_{0}$,
 eq. (\ref{diag1}), and for unequal quark masses, the hamiltonian is
linear in the quadratic Casimir operator. To see this we now
derive the hamiltonian corresponding to the action (\ref{lag2}).
The canonical momenta are given by \br\label{canonical} p_{\a} =
\frac{\pa\,\,\, L}{\pa \dot{z}_{\a}^{(N_{f})\, \star}}=
\frac{1}{2M_{N_{f}}}\[\dot{z}_{\a}^{(N_{f})}-
\(\dot{z}_{\b}^{(N_{f})} z_{\b}^{(N_{f})\, \star}\)
z_{\a}^{(N_{f})}\] + \frac{iN_{C}}{2} z_{\a}^{N_{f}} \er and there
is a conjugate expression for $ p_{\a}$. Therefore, from $ H =
p_{\a} \dot{z}_{\b}^{(N_{f})\,\star} + p_{\a}^{\star}
\dot{z}_{\b}^{(N_{f})} -L $,\, one can get the hamiltonian \br
\label{hamiltonian0} H = \frac{1}{2M_{N_{f}}}
\(Dz\)^{\dagger}_{\a} \( D z\)_{\a} + \frac{2\pi}{M_{N_{f}} N_{C}}
\sum_{i=1}^{N_{f}} \(
\widetilde{m}^{2}_{i}-\widetilde{m}^{2}_{N_{f}}\) z_{i}^{(N_{f})\,
\star} z_{i}^{(N_{f})}. \er

However, one must take a proper care of the relevant constraint
(\ref{compl2}) which was incorporated through the addition of a
Lagrange multiplier in the action (\ref{lag2}). A proper treatment
of a constrained system must be performed at this point
\cite{frishman}. In \cite{frishman, frishman87} it was shown that
the local $U(1)$ gauge symmetry (\ref{u1}) leads to the constraint
\br \label{u1const} Q_{N_{f}\, N_{f}}=0\,\, \Rightarrow\,\, {\cal
Q}_{B} = \sqrt{2N_{f} (N_{f}-1)}\, Q_{Y} \er which has to be
imposed on physical states. This implies that  the representation
$R$ must contain a state with $Y$ charge \br \bar{Q}_{Y} =
\sqrt{\frac{1}{2N_{f}(N_{f}-1)}}\, N_{C}. \er

The remaining states will be generated through the application of
the $SU(N_{f})$ transformations to this one. For states with only
quarks and no antiquarks, the condition that ${\cal Q}_{B} =
N_{C}$ implies that only representations described by Young
tableaux with $N_{C}$ boxes appear. The additional constraint that
$Q_{Y} = \bar{Q}_{Y}$ implies that all $N_{C}$ quarks belong to
$SU(N_{f}-1)$, i.e., this state does not involve the $N_{f}^{\,
'th}$ quark flavor. These constraints are automatically satisfied
in the totally symmetric representation of $N_{C}$ boxes, which is
the only representation possible in two dimensions. This is
because the state wave functions have to be constructed out of the
components of the complex vector $z^{(N_{f})}$ as\, \br
\psi(z^{(N_{f})}\,,z^{(N_{f})\, \star}) =
(z_{1}^{(N_{f})})^{p_{1}}...(z_{N_{f}}^{(N_{f})})^{p_{N_{f}}}(z_{1}^{(N_{f})\,
\star})^{q_{1}}...(z_{N_{f}}^{(N_{f})\,\star})^{q_{N_{f}}}\label{multiplet1}\er
with $\sum_{i=1}^{N_{f}} (p_{i}-q_{i})=N_{C}.$

The lowest such multiplet has \br \sum_{i=1}^{N_{f}} p_{i}=N_{C}
\,\,\,\, \mbox{and all}\,\,\,\, q_{i}=0 \label{multiplet2}\er.

This multiplet corresponds to the Young
tableaux\begin{eqnarray}\label{young} \overbrace{\Box \cdots
\Box}^{N_c}
\end{eqnarray}

In QCD$_{2}$ for $N_{C} = 3,\,N_{F} = 3$\, we get only the ${\bf
10}$ of $SU(3)$.

Then, taking into account (\ref{casimir2}), (\ref{casimir00}) and
(\ref{masspar}), the expression (\ref{hamiltonian0}) becomes
\cite{frishman, ellis1} \br \label{hamiltonian1} H =
M_{baryon}^{cl} \Big\{ 1+ \(\frac{\pi}{2N_{C}}\)^2 \[ C_{2}(R) -
\frac{n^2_{N_{f}} N_{C}^2}{2N_{f}} (N_{f}-1)\] +
\sum_{i=1}^{N_{f}} \frac{m_{i}-m_{N_{f}}}{m_{N_{f}}}
|z_{i}^{(N_{f})}|^2 \Big\}, \er where $M_{baryon}^{cl}$ is given
by (\ref{baryonmass}) and $C_{2}(R)$ is the value of the quadratic
Casimir  for the flavor representation $R$ of the baryon. For a
baryon state given by SG
 1-soliton solution one must set $n_{N_{f}}=1$ in the hamiltonian above.
 Notice that the Hamiltonian depends on $m_{0}$ only through $M_{baryon}^{cl}$, so the overall
mass scale is undetermined, only the mass ratios are meaningful.
The mass term contributions come from quantum fluctuations around
the classical soliton, consistency with the semi-classical
approximation requires that it be very small compared to one.
However, these terms vanish for equal  quark masses
\cite{frishman, ellis}. The ${\bf 10}$ baryon mass becomes

\br \label{qbaryonmass} M(baryon) &=& M_{classical} \[ 1+
 (\frac{\pi^2}{8}) \frac{N_{f}-1}{N_{C}} \].\er

Notice that the quantum correction is suppressed by a factor of
$N_{C}$. Moreover, the quantum correction for $N_{C}=3,\,N_{f}=3$
numerically becomes $\sim 0.82$.

The hamiltonian (\ref{hamiltonian1}) taken for equal quark masses
 has been used to compute the energy of the first exotic baryon
${\cal E}_{1}$(a state containing $N_{C}+1$ quarks and just one
anti-quark) by taking the corresponding Casimir $C_{2}({\cal
E}_{1})$ for $R=\bf{35}$ of flavor relevant to the exotic state
\cite{ellis}. For further analysis we record the mass of this
exotic baryon \br M({\cal E}_{1}) = M(classical) \[1+
\frac{\pi^2}{8}\frac{1}{N_{C}} \(3+ N_{f} -\frac{6}{N_{f}}\) +
\frac{3\pi^2}{8} \frac{1}{N_{C}^2} \( N_{f} - \frac{3}{N_{f}}\)
\].\label{exotic0}\er
In the interesting case $N_{C}=3,\,N_{f}=3$ this becomes \br
M({\bf 35}) = M(classical) \Big\{1+
\frac{\pi^2}{4}\Big\}.\label{exotic01}\er

In this case the correction due to quantum fluctuations around the
classical solution is larger than the classical term. So, the
 semi-classical approximation may not be a good approximation.
 However, observe that the ratio $M({\bf 35}) /M({\bf 10}) \sim 1.9 $, which
 is $17\%$ larger than the ratio between the experimental masses
 of the $\Theta^{+}$ and the nucleon. See more on this point
 below. These semi-classical approximations may be improved by introducing
 different ansatz for $g_{0}$ and considering unequal quark mass
 parameters. These points will be tackled in the next sections.

\section{The GSG model, the unequal quark
masses and baryon states} \label{gsgqcd}

 In the following we will
concentrate on the effective action (\ref{lag1}) for the
particular case $N_{f}=3$ and unequal quark mass parameters. So,
the $SU(3)$ flavor symmetry is broken explicitly by the mass
terms.

The effective Lagrangian in the case of $N_{f}=3$ from
(\ref{lag1}), upon using (\ref{compl2}), can be written as
\br\nonumber
 S[z_{p}^{(i)}(t)]&=&
  \frac{1}{4} \int dt \,\Big\{ \(M_{12}^{-1}+ M_{13}^{-1}-M_{23}^{-1} \) \[ \dot{z}^{(1)}_{\a} \dot{z}^{(1)\,\star}_{\a}
 - \dot{z}^{(1)}_{\a} z^{(1)\,\star}_{\a} z^{(1)}_{\b}
 \dot{z}^{(1)\,\star}_{\b}\]+\\&&
 \(M_{12}^{-1}- M_{13}^{-1}+M_{23}^{-1} \) \[ \dot{z}^{(2)}_{\a} \dot{z}^{(2)\,\star}_{\a}
 - \dot{z}^{(2)}_{\a} z^{(2)\,\star}_{\a} z^{(2)}_{\b}
 \dot{z}^{(2)\,\star}_{\b}\]+
 \nonumber\\&&
 \(-M_{12}^{-1}+ M_{13}^{-1}+M_{23}^{-1} \) \[ \dot{z}^{(3)}_{\a} \dot{z}^{(3)\,\star}_{\a}
 - \dot{z}^{(3)}_{\a} z^{(3)\,\star}_{\a} z^{(3)}_{\b}
 \dot{z}^{(3)\,\star}_{\b}\]
\Big\}
 -\nonumber\\
&& i \frac{N_{C}}{2} \int dt  \sum_{i, p} n_{i} \[
\dot{z}_{p}^{(i)} z^{(i)\,\star}_{p}-  z_{p}^{(i)}
\dot{z}^{(i)\,\star}_{p}\]- \nonumber\\&& \int dt
\Big\{\frac{2\pi}{N_{c}} \sum_{i,\,p}
\[ \frac{\widetilde{m}^2_{p}}{M_{i}} - \frac{\widetilde{m}^2_{k}}{M_{k}}\] z_{p}^{(i)}
z_{p}^{(i)\,\star} + \frac{2\pi}{N_{c}}
\[ \sum_{i} \frac{\widetilde{m}^2_{i}}{M_{i}} - \frac{\widetilde{m}^2_{k}}{M_{k}} N_{\phi}
\]\Big\}
\label{lag33} \er

From (\ref{casimir1}) and following similar steps the second
$U(3)$ Casimir operator can be written as\br Q^{A} Q_{A}&=&
\frac{1}{2} Q_{\a\b}Q_{\b \a},\nonumber\\\nonumber &=&
\frac{1}{2}N_{C}^2 \sum_{j} n_{j}n_{j} + \frac{1}{8} \Big\{
\(M_{12}^{-2}+ M_{13}^{-2}-M_{23}^{-2} \) \[ \dot{z}^{(1)}_{\a}
\dot{z}^{(1)\,\star}_{\a}
 - \dot{z}^{(1)}_{\a} z^{(1)\,\star}_{\a} z^{(1)}_{\b}
 \dot{z}^{(1)\,\star}_{\b}\]+\\&&
 \(M_{12}^{-2}- M_{13}^{-2}+M_{23}^{-2} \) \[ \dot{z}^{(2)}_{\a} \dot{z}^{(2)\,\star}_{\a}
 - \dot{z}^{(2)}_{\a} z^{(2)\,\star}_{\a} z^{(2)}_{\b}
 \dot{z}^{(2)\,\star}_{\b}\]+
 \nonumber\\&&
 \(-M_{12}^{-2}+ M_{13}^{-2}+M_{23}^{-2} \) \[ \dot{z}^{(3)}_{\a} \dot{z}^{(3)\,\star}_{\a}
 - \dot{z}^{(3)}_{\a} z^{(3)\,\star}_{\a} z^{(3)}_{\b}
 \dot{z}^{(3)\,\star}_{\b}\]
\Big\} \label{casimir3}.\er

As a particular case for the ansatz (\ref{diag1}) let us take
$N_{f}=3$, so $n_{1}=n_{2}=0$\, in (\ref{indices}). In (\ref{mij})
one can set formally $M_{12}\equiv +\infty$\,\,and in view of
  (\ref{mm1})  the remaining parameters can be written as\, $M_{13}=M_{23}\equiv
 M_{3}$. Thus, taking into account these parameters the expressions for the action (\ref{lag33}) and the
 second Casimir (\ref{casimir3}) reduce to the well known ones
(\ref{lag2}) and (\ref{casimir2}), respectively.

Next, we discuss the action (\ref{lag33}) and the
 second Casimir (\ref{casimir3}) operator for the soliton and kink type solutions of the GSG model. In appendices
 \ref{s11}, \ref{s22} and \ref{s33} we classify these type of solutions. There are three $1-$soliton solutions
 [see eqs. (\ref{sol1}), (\ref{sol2}) and (\ref{sol3b})] which correspond to baryon number $N_{C}$
 [see eqs. (\ref{charge1}), (\ref{charge2}) and (\ref{charge3})], because the GSG model
possesses the
 symmetry (\ref{symm}) the third soliton is doubly
 degenerated. From the fields relationships (\ref{relation1}),
(\ref{relation2}) and
 (\ref{relation3}) one has the three $1-$soliton cases
\br i)&& \,\,\,\,\Phi_{1}=-\Phi_{2}=\Phi_{3}=\vp_{1}\,\,\,\,\,\,
\Rightarrow\,\,
 M_{13}=+\infty,\,\, M_{12}=M_{23} \equiv {\cal M}_{2};\,\,\,\,M_{1}=M_{2}=M_{3}=\widetilde{M}_{2},\nonumber \\
\label{mcal2}\\ ii)&& -\Phi_{1}=\Phi_{2}=\Phi_{3}=\vp_{2}
\,\,\,\,\,\,\,\,\,\,
 \Rightarrow\,\,\,
M_{23}=+\infty,\,\, M_{12}=M_{13} \equiv {\cal
M}_{1};\,\,\,\,M_{1}=M_{2}=M_{3}=\widetilde{M}_{1},\nonumber \\\label{mcal1} \\
 iii)&& \,\,\,\,\Phi_{1}=\Phi_{2}=-\Phi_{3}=\hat{\vp}
\,\,\,\,\,\,\,\,\, \Rightarrow\,\, M_{12}=+\infty,\,\,
M_{13}=M_{23} \equiv {\cal M}_{3}
;\,\,\,\,M_{1}=M_{2}=M_{3}=\widetilde{M}_{3}\nonumber \\
\label{mcal3}\er where the eqs. (\ref{mij}) and (\ref{mm1}) have
been used, respectively, to define the parameters ${\cal M}_{j}$
and $\widetilde{M}_{j}$ in the right hand sides of the
relationships above.

In appendix \ref{dsg:sec} we record the kink type solution [see
eq. (\ref{kink})] which corresponds to the GSG reduced model
called double sine-Gordon theory. This solution corresponds to
baryon number $4N_{C}$ [see eq. (\ref{chargekink})]. Thus, from
(\ref{fieldskk1}), (\ref{mij}) and (\ref{mm1}) one has
\br\nonumber  \Phi_{1}=\Phi_{2}=\frac{1}{2} \Phi_{3}=\frac{1}{2}
\phi \,\, \Rightarrow\,\, M_{12}=+\infty,\,\, M_{13}=M_{23} \equiv
 {\cal M}_{K};\,\,\,\,M_{1}=M_{2}\equiv {\cal M}_{K},\,\,M_{3}\equiv {\cal M}_{2K}\\ \label{mkink}\er

The solutions with baryon numbers $2N_{C}$ and $3N_{C}$ correspond
to composite configurations formed by multi-solitons of the GSG
model. These states (i.e. multi-baryons) deserve a careful
treatment which we hope to undertake in future.

\subsection{GSG solitons and the states with baryon number $N_{C}$}
\label{nc}

For the particular cases (\ref{mcal2})-(\ref{mcal3}) one can
rewrite the action (\ref{lag33}) such that for each case the terms
quadratic in time derivatives reduce to a term depending only on
one variable, say $z^{(\hat{l})}_{i}$, related to the $\hat{l}$'th
column of the matrix $A$. The reason is that the symmetries of the
quantum mechanical lagrangian and actual manifold on which $A(t)$
lives depend on the properties of the ansatz $g_{0}$. For the
ansatz $g_{0}$ related to the GSG model one can see that the
space-time dependent field $g$ in eq. (\ref{rota1}) can be
rewritten only in terms of certain columns of $A$. For example, in
the case (\ref{mcal3}) above the matrix $g(x,
t)$ can be written as  \br g_{\a\b} (x,t)&=& [A g_{0} A^{-1}]_{\a\b} \nonumber\\
&=& \d_{\a\b} e^{i\b_{0} \hat{\vp}}- 2 i \, \mbox{sin} (\b_{0}
\hat{\vp})z_{\a}^{(3)} z_{\b}^{(3)\, \star}, \er which clearly
depends only on the third column of $A$. So, we may think the left
hand side of (\ref{g0}), i.e. \, $[\widetilde{S}(g(x,
t))-\widetilde{S}(g_{0}(x))]$, entering the expression of the
semi-classical quantization approach, would in principle be
written only in terms of the third column of $A$. However, in
order to envisage certain local symmetries it is useful to write
the terms first order in time derivatives as depending on the full
parameters $z_{i}^{(j)}$ of the field $A$. These terms arise from
the WZW term and provides the Gauss law type $N_{z}$ number
conservation law [See eq. (\ref{numbernz}) below]. An additional
$SU(2) \in H $ (see (\ref{subgroup})) local symmetry will be
described below. Moreover, this picture is in accordance with the
counting of the degrees of freedom. In fact, the effective action
(\ref{g0}) possesses the local gauge symmetry
 (\ref{symm11}), where in the case of field configuration (\ref{mcal3}) the gauge group
 $H$ becomes
 \br
 \label{subgroup}
 H = SU(2) \times U(1)_{B} \times U(1)_{Y},
 \er
with the last two $U(1)$ factors related to baryon number and
hypercharge, respectively. Thus, the effective action
(\ref{lag33}) will be an action for the coordination describing
the coset space $G/H = SU(3) \times U(1)_{B} / SU(2) \times
U(1)_{B} \times U(1)_{Y}\, =\, CP^2$. The $\Phi_{i}$ fields and
symmetries of $g_{0}$ also determine the values and relationships
between the parameters $M_{ij}$ in (\ref{mcal2})-(\ref{mcal3}),
such that certain coefficients in (\ref{lag33}) depending on these
parameters vanish identically, thus leaving
 a subset of $z_{i}^{(j)}$ variables which
 must be consistent with the counting of the degrees of freedom.
 For example this picture is illustrated in the case (\ref{mcal3})
 where the coefficients $(M_{12}^{-1}+ M_{13}^{-1}-M_{23}^{-1} )$ \, and $(M_{12}^{-1}- M_{13}^{-1}+M_{23}^{-1}
 )$ vanish identically, leaving an action with kinetic term depending only
 on the variables $z_{\a}^{(3)}$. However, the mass and WZW
 terms are conveniently written in terms of the complete
 $z_{i}^{(j)}$ variables.

So, for each case in (\ref{mcal2})-(\ref{mcal3}) labelled by
$\hat{l}$, the action can be written as \br\nonumber
 S[z_{p}^{(i)}(t)]&=&
  \frac{1}{2} \int dt \, {\cal M}_{\hat{l}}^{-1} \[ \dot{z}^{(\hat{l})}_{\a} \dot{z}^{(\hat{l})\,\star}_{\a}
 - \dot{z}^{(\hat{l})}_{\a} z^{(\hat{l})\,\star}_{\a} z^{(\hat{l})}_{\b}
 \dot{z}^{(\hat{l})\,\star}_{\b}\]-\nonumber\\&&
i \frac{N_{C}}{2} \int dt  \sum_{i,\,p} n_{i} \[ \dot{z}_{p}^{(i)}
z^{(i)\,\star}_{p}-  z_{p}^{(i)} \dot{z}^{(i)\,\star}_{p}\]
 - \frac{2\pi}{N_{c}\widetilde{M}_{\hat{l}}} \int dt \sum_{i,\, j} \widetilde{m}^2_{i} |z_{i}^{(j)}|^2  \label{lag44}. \er

In the relation above we must assign the relevant set of values to
the indices $n_{i}\,(i=1,2,3)$ (see Appendix) for the relevant
case in (\ref{mcal2})-(\ref{mcal3}). The first term in
(\ref{lag44}) is the usual CP$^{2}$ quantum mechanical action,
while the terms first order in time-derivatives are modifications
due to the WZ term, as arisen from (\ref{g0}) and (\ref{lag1}).
Notice that the last term was originated from the unequal quark
mass terms.

Following similar steps as in the single baryon case (see eqs.
(\ref{canonical})-(\ref{hamiltonian0})) one can obtain the
hamiltonian \br \label{hamiltonian11} H &=& \frac{1}{2{\cal
M}_{\hat{l}}} \(Dz^{(\hat{l})}\)^{\dagger}_{\a} \( D
z^{(\hat{l})}\)_{\a} +
\frac{2\pi}{N_{c}\widetilde{M}_{\hat{l}}}\sum_{i,\,j}
\widetilde{m}^2_{i} |z_{i}^{(j)}|^2 ,\er where $ \( D
z^{(\hat{l})}\)_{\a}= \dot{z}_{\a}^{(\hat{l})} -
z_{\a}^{(\hat{l})} (z_{\b}^{(\hat{l})\, \star}
\dot{z}_{\b}^{(\hat{l})}).$

Similarly, the corresponding second Casimir becomes \br Q^{A}
Q_{A}&=& \frac{1}{2} Q_{\a\b}Q_{\b \a},\nonumber\\ &=&
\frac{1}{2}N_{C}^2 \sum_{i} |n_{i}|^2 + \frac{1}{4 {\cal
M}_{\hat{l}}^{2}} \(Dz^{(\hat{l})}\)^{\dagger}_{\a} \( D
z^{(\hat{l})}\)_{\a} \label{casimir44}\er

Then from (\ref{hamiltonian11})-(\ref{casimir44}) and taking into
account $Q^{A} Q_{A}= C_{2} + \frac{1}{2N_{f}} \sum_{i} ({\cal
Q}_{B}^{i})^2$\, one can get \br \label{hamiltonian22} H &=& 2
{\cal M}_{\hat{l}} \( C_{2}+ \frac{1}{2N_{f}} \sum_{i} ({\cal
Q}_{B}^{i})^2 - \frac{1}{2} N_{C}^2 \sum_{i} |n_{i}|^2\)+
\frac{2\pi}{N_{c}\widetilde{M}_{\hat{l}}}\sum_{i,j=1}^3
\widetilde{m}^2_{i}|z_{i}^{(j)}|^2,\er where $ {\cal Q}_{B}^{i}=
n_{i} N_{C}$ for a convenient choice of the indices $n_{i}$, which
in the cases (\ref{mcal2})-(\ref{mcal3}) is simply $|n_{i}|=1$
[see also eqs. (\ref{charge1}), (\ref{charge2}) and
(\ref{charge3}) for 1-soliton configurations]. The parameters
${\cal M}_{\hat{l}},\, \widetilde{M}_{\hat{l}}$ can be computed
for the relevant solitons. They become \br \frac{1}{2{\cal
M}_{\hat{l}}} = \frac{1}{\widetilde{m}} \frac{2\sqrt{2}}{3}
(\frac{N_{C}}{\pi})^{3/2},\,\,\,\,\,\,\,\,\,\frac{1}{2\widetilde{M}_{\hat{l}}}
= \frac{1}{\sqrt{2}\,\widetilde{m}} (\frac{N_{C}}{\pi})^{3/2}
\label{mm2}\er

Some comments concerning the two hamiltonians
 (\ref{hamiltonian1}) and (\ref{hamiltonian22}) are in order here.
Even though they correspond to one baryon state (baryon number
$N_{C}$) they look different. In fact, the hamiltonian
(\ref{hamiltonian22}) incorporates additional terms. First, due to
the ansatz (\ref{diag0}) related to the GSG model one has some set
of field solutions comprising in total three possibilities
(\ref{mcal2})-(\ref{mcal3}) with baryon number $N_{C}$, each case
being characterized by the set of parameters ${\cal
M}_{\hat{l}},\, \widetilde{M}_{\hat{l}}$  and relevant
combinations of the indices $n_{j}$ which are related to the
baryon number of the configuration $\{{\Phi_{j}} \},\,j=1,2,3$.
So, the terms $-\frac{N_{C}^2}{2}$\, and\,
$\frac{N_{C}^2}{2N_{f}}$ in (\ref{hamiltonian1}) translate to
$-\frac{N_{C}^2}{2}\sum_{i} n_{i}^2$ \,and\, $\frac{1}{2N_{f}}
\sum_{i} ({\cal Q}_{B}^{i})^2$, respectively, in the new
hamiltonian (\ref{hamiltonian22}). Second, the mass term
expression allows an exact summation due to unitarity, thus giving
a constant additional term to the hamiltonian (see below). The
corresponding term in (\ref{hamiltonian1}), obtained in
\cite{ellis1}, does not permit an exact summation.

\subsection{Lowest lying baryon state and the GSG soliton}
\label{lowest}

So far, the treatment for each case (\ref{mcal2})-(\ref{mcal3})
followed similar steps; however, in order to compute the quantum
correction to the soliton mass we choose the one from the
classification  (\ref{mcal2})-(\ref{mcal3}) with the minimum
classical energy solution. Thus, taking into account the
``physically" motivated inequalities $m_{3}< m_{1}< m_{2}$ ( or
\,$ \m_{3}< \m_{1}< \m_{2}$) [eq. (\ref{mu11}) relates the
$\mu_{j}$'s and the $m_{j}$'s] one observes that the soliton with
mass $M_{2}^{sol}$\, [see eq. (\ref{solmass2})] possesses the
smallest mass according to the relationship (\ref{relat12}). This
corresponds to the {\sl second case} (\ref{mcal1}) classified
above; so one must set the index $\hat{l}=1$ in the action
(\ref{lag44}).

The variation of the action  (\ref{lag44}) under $z_{\a}^{(j)}
\rightarrow e^{i\d(t)} z_{\a}^{(j)}$ is due to the WZW term: $\D S
= N_{c}(n_{1}+ n_{2}+ n_{3}) \int dt\, \dot{\d}$. This implies
\br\label{numbernz} N_{z} = \frac{\D S}{\D \dot{\d}} = N_{c}\(
n_{1}+n_{2}+ n_{3}\),\er which is an analog of the Gauss law, and
restricts
 the allowed physical states \cite{rabinovici}. For the soliton configuration with baryon number $N_{C}$,\,
 (\ref{mcal1}), under consideration in this subsection, we have
 $n_{1}=-n_{2}=-n_{3}=-1$ $\rightarrow$ $n_{1}+n_{2}+n_{3}=1$ \,[see eq. (\ref{nn2}) ]\, implying \, \br N_{z} =
 N_{C}\label{numberz1}.\er

Therefore, for any wave function, written as a polynomial in $z$
\, and \, $z^{\star}$\, the number of the  $z$ minus the number of
the  $z^{\star}$ must be equal to $N_{C}$. But due to a larger
local symmetry we will have more restrictions. Thus, as commented
earlier the (massless part) effective action (\ref{lag44}) is
invariant under the local $SU(2)$ symmetry. This can be easily
seen by defining ``local gauge potentials" \br
\widetilde{A}_{\b\a}(t)= - \sum_{p} z^{(\b)\,\star}_{p}
\dot{z}^{(\a)}_{p},\,\,\,\,\a , \b=2,3. \er

Under the local gauge transformation corresponding to $\L(t)$, one
has \br\label{u2} \widetilde{A}(t) \rightarrow
e^{i\L}\widetilde{A} e^{-i\L} + \pa_{t} e^{i\L} e^{-i\L}.\er

Then we have that the WZW term in (\ref{lag44}) for the variables
$z_{p}^{\a},\,\,\a=2,3$ (take $\hat{l}=1$, \, $n_{2}=n_{3}=1$)
remain invariant under the transformation (\ref{u2})

\br iN_{C} \int dt\, \mbox{Tr}\, \dot{z}^{(\a)\,\star}_{p}
z^{(\b)}_{p} \equiv iN_{C}\int dt \,\mbox{Tr}\, \widetilde{A}
\,\,\Rightarrow\,\, iN_{C} \int dt\, \mbox{Tr} \widetilde{A} \er

Remember that the variables $z_{p}^{\a}$ do not appear in the
kinetic term of (\ref{lag44}). The local symmetry above imply that
the allowed physical states must be singlets under the $SU(2)$
symmetry in flavor space. So, the wave functions for $z's$ only
(analogous to quarks only for QCD) must be of the form \br
\psi_{2}(z) = \Pi_{i=1}^{N_{C}} \(\epsilon_{\a_{1}\a_{2}}
\,z_{i_{1}}^{(\a_{1})}z_{i_{2}}^{(\a_{2})}\),\,\,\,\,\a_{1},\a_{2}
= 2, 3  \label{confined1},\er where $1\leq i_{1}, i_{2}\leq
N_{f}.$

Then, taking into account the restrictions of the types
(\ref{numberz1}) and (\ref{confined1}) the most general state can
be written as \br \widetilde{\psi}(z, z^{\star}) = \psi_{2}(z)\[
\Pi_{\{p,q\}} (z^{(\a)\, \star}_{p} z_{q}^{(\a)})^{n_{pq}}\], \er
and the products are defined for some sets of indices. This wave
function generalizes the one given in (\ref{multiplet1}).

Next, let us compute the mass of the state represented for wave
functions of the form $\widetilde{\psi}(t) = \psi_{2}(z) \,
\Pi_{i}(z_{i}^{(1)})^{p_{i}}$ \,\,where\, ($\sum_{i=1}^{N_{f}}
p_{i}=N_{C}$).

Combining the hamiltonian (\ref{hamiltonian22}), the relevant
parameters (\ref{mm2}) and the classical soliton mass term, for
the $R$ baryon we have \br M(baryon) &=& M_{classical} \Big\{
1+\frac{3}{4} (\frac{\pi}{2N_{C}})^2\[ C_{2}(R) -
\frac{N_{C}^2}{2} (N_{f}-1) + \frac{1}{2 \widetilde{m}^2} \sum_{i}
\widetilde{m}_{i}^{2}
\]\Big\} \label{qbaryonmass1}\er
where \br M_{classical}=4 \widetilde{m}
(\frac{2N_{C}}{\pi})^{1/2},\,\,\,\,\,\,\widetilde{m}^2 =
\frac{1}{13} (\frac{m^2}{m_{0}}) \( 6 m_{1} + 3 m_{2} \).\er

The last term in (\ref{hamiltonian22}) simplifies to a constant
term by unitarity condition of the matrix elements $z_{i}^{(j)}$
and the parameter $\widetilde{m}$ corresponds to the one-soliton
parameter once the identification $\g^{2}_{2} = 2 \b_{0}^2
\widetilde{m}^2$ is made in (\ref{gamma2})  by comparing the SG
one-solitons (\ref{soliton}) and (\ref{sol2}). Even though the
computations are explicitly made for $N_{f}=3$ it is instructive
to leave the number of flavors as a variable. In the case of the
${\bf 10}$ baryon one has \br \label{qbaryonmass11}M(baryon) &=&
M_{classical}
\[ 1+\frac{3\pi^2}{32} \frac{N_{f}-1}{N_{C}} - \frac{3 \pi^2}{32} \frac{(N_{f}-1)^2}{N_{f}} + \frac{3}{2} \].\er

In the following we discuss the correction terms to the earlier
expression (\ref{qbaryonmass}) for the ${\bf 10}$ baryon as
compared to the last improved expression (\ref{qbaryonmass11}).
The quantum correction of (\ref{qbaryonmass}) is multiplied by
$3/4$ and the last two terms in (\ref{qbaryonmass11})
 are new contributions due to the GSG ansatz used and the unequal quark mass terms. The last term contribution in
 (\ref{qbaryonmass1}) was simplified providing a numerical
 term $3/2$ in (\ref{qbaryonmass11}) thanks to unitarity and the relationship
 between the quark masses (\ref{gamma2}) which is a condition to
 get the relevant soliton solution. This term apparently may not
 be consistent with a quantum correction around the classical solution since consistency with the semi-classical
 approximation requires it be small compared to one. However,
 this term must be combined with the third term which gives a negative value contribution and is
 an additional term independent of $N_{C}$, as is the last numerical
 $3/2$ term under discussion. In fact, for $N_{C}=3,\,N_{f}=3$, numerically these
 two terms contribute $\sim 0.27$, which is acceptable. The $N_{C}$
 dependent term numerically becomes $\sim 0.62$ (the term $0.82$ of (\ref{qbaryonmass}) has been multiplied by
 $3/4$). Adding all the quantum contributions one has $0.89$, which
 increases the earlier numerical value $0.86$ of (\ref{qbaryonmass}) in  $\sim 3.5\%$. In fact, this is a small correction to the already known value
which was obtained using the ansatz
 (\ref{diag1}) in \cite{frishman, ellis}.

\subsection{Possible vibrational modes and the GSG model}
\label{vibr}

The only static soliton configurations with baryon number $N_{C}$,
which emerge in the strong-coupling regime of QCD$_{2}$, are the
ones we have considered above in eqs. (\ref{mcal2})-(\ref{mcal3}).
Precisely, these are the one-solitons of the GSG model which, in
subsection \ref{lowest}, have been the subject of semi-classical
treatment. Their quantum corrections by time-dependent rotations
in flavor space have been computed, we focused on the one with the
lowest classical mass. Since in two dimensions there are no spin
degrees of freedom, in order to search for higher excitations we
must look for vibrational modes which might in principle exist.
These type of excitations in the strong coupling limit can be
found as classical time-dependent solutions of the GSG equations
of motion (\ref{eq1})-(\ref{eq2}). Looking at time-dependent
solutions of type (\ref{mcal1}) [see eq. \ref{sol2}] one has that
the field $\vp_{2}$ satisfies ordinary sine-Gordon equation \br
\pa_{tt} \vp_{2} -\pa_{xx} \vp_{2} + 2 \widetilde{m}^2
\sqrt{\frac{4\pi}{N_{C}}} \mbox{sin}\(\frac{4\pi}{N_{C}}
\vp_{2}\),\,\,\,\,\,\,\vp_{1}(x,t) \equiv 0.\label{sgt}\er

The time dependent one-soliton solution of (\ref{sgt}) for the
field $\vp_{2}$, determines the configuration
$\{\Phi_{1},\Phi_{2}, \Phi_{3}\}$ in (\ref{mcal1}) with baryon
number $N_{C}$ in the QCD$_{2}$ context. To look for higher
excitations, for example, one can search for a coupled state of
one-baryon and {\sl breather} type vibrations (soliton-antisoliton
bound states) of the GSG system, which can give a total baryon
number $N_{C}$. We were not be able to find a more general
time-dependent mixed single-baryon plus vibrational state with
baryon number $N_{C}$ for the general GSG equation. For example,
this type of solution, if it exists, may be useful in order to
study meson-baryon scattering as considered in \cite{ellis2}. As
it is well known the SG eq. (\ref{sgt}) does give vibrational
solutions in the form of breather states (meson states), for later
use we simply recall that in the large $N_{C}$ limit the
lowest-lying mesons have masses of order $\sqrt{m_{q} e_{c}}$
\cite{Rajaraman} ($m_{q}$ is defined in eq. (\ref{delta}) below).
We refer the reader to ref. \cite{ellis} for more discussion, such
as the various meson couplings to baryons with different degrees
of exoticity.

\section{The GSG solitons and the exotic baryons}
\label{exo}
\subsection{The first exotic baryon}

Here we will follow the analog of the rigid-rotor approach (RRA)
to quantize solitons and obtain exotic states. In this method it
is assumed that the higher order representation multiplets are
different rotational (in spin and isospin) states of the same
object (the ``classical baryon", i.e the soliton field)
\cite{Diakonov1}. This assumption has allowed in the past the
obtention of some relations between the characteristics of the
nonexotic baryon multiplets which are satisfied up to a few
percent in nature. However, see refs. \cite{cohen, klebanov} for
some critiques to this conventional approach for exotic baryons.
According to these authors the conventional RRA, in which the
collective rotational approach and vibrational modes of the
soliton are assumed to be decoupled, and only the rotational modes
are quantized, is only justified  at large $N_{C}$ for nonexotic
collective states in $SU(3)$ models. On the other hand, the bound
state approach (BSA) to quantize solitons, due to Callan-Klebanov
\cite{klebanov}, considers broken $SU(3)$ symmetry in which the
excitations carrying strangeness are taken as vibrational modes,
and should be quantized as harmonic vibrations. However, for
exotic states the Callan-Klebanov approach does not reproduce the
RRA result; indeed this approach gives no exotic resonant states
when applied to the original Skyrme model \cite{klebanov}. There
was intensive discussion of connections between the both
 approaches mentioned above. The rotation-vibration approach
(RVA) (see \cite{walliser} and references therein) includes both
rotational (zero modes) and vibrational degrees of freedom of
solitons and is a generalization of the both methods above, which
therefore appear in some regions of the RVA method when certain
degrees of freedom are frozen. A major result of the RVA method is
that pentaquark states {\sl do} indeed emerge in both methods
above, i.e. in the RRA and BSA. In order to illustrate the present
situation of the theoretical controversy let us mention that the
RVA approach was criticized in \cite{cohen1}, and the reply to
this criticism was given in \cite{walliser1}.

Following the analog of the RRA, the expression
(\ref{qbaryonmass1}) can be used to compute the energy of the
first exotic baryon ${\cal E}_{1}$ (a state containing $N_{C}+1$
quarks and one antiquark) by taking the corresponding Casimir
$C_{2}({\cal E}_{1})$ for $R=\bf{35}$ of flavor relevant to the
exotic state in two-dimensions. This state is an analogue of the
${\bf \overline{10}}, {\bf 27}$ and ${\bf 35}$ states in four
dimensions. So,  following \cite{ellis}, in the conventional RRA
one has that the mass of the first exotic state becomes \br
M({\cal E}_{1}) = M(classical) \Big\{1+ \frac{3}{4}
\[\frac{\pi^2}{8}\frac{1}{N_{C}} \(3+ N_{f} -\frac{6}{N_{f}}\) + \frac{3\pi^2}{8}
\frac{1}{N_{C}^2} \( N_{f} - \frac{3}{N_{f}}\)\]\nonumber \\
-\frac{3\pi^2}{32} \frac{(N_{f}-1)^2}{N_{f}} + \frac{3}{2}
\Big\}\label{exotic1}\er

In the interesting case $N_{C}=3,\,N_{f}=3$ this becomes
\br\label{exotic11} M({\bf 35}) =M(classical) \Big\{ 1+
\frac{3}{4} \frac{\pi^2}{4} -\frac{\pi^2}{8} +
\frac{3}{2}\Big\}.\er

In this case the correction due to quantum fluctuations around the
classical solution is still larger than the classical term, as it
was in the earlier computation (\ref{exotic01}). However,
numerically in eq. (\ref{exotic11}) the correction is $2.12$,
whereas in eq. (\ref{exotic01}) it was $2.46$. In fact, the
contribution in (\ref{exotic11}) decreases in $0.34$ units the
earlier computation. So, we may claim that the introduction of
unequal quark masses and the ansatz given by the GSG model
slightly improve the semi-classical approximation.

Moreover, notice that the ratio of the experimental masses of the
$\Theta^{+}(1530)$ and the nucleon is $1.63$. On the other hand,
the ratio of the first exotic to that of the lightest baryon in
the QCD$_{2}$ model becomes \br\label{fracao} \frac{M_{{\bf
35}}}{M_{10}} = \frac{1+ \frac{3\pi^2}{16} -\frac{\pi^2}{8} +
\frac{3}{2}}{1+\frac{\pi^2}{16}-\frac{\pi^2}{8} + \frac{3}{2}}
\sim  1.65,\er which is only $1\%$ larger to its 4D analog. This
must be compared to the earlier calculation which gave a value
$17\%$ larger [see eq. (\ref{exotic01})]. However, the result in
(\ref{fracao}) could be a numerical coincidence, since in two
dimensions we are not considering the spin degrees of freedom that
is important in QCD$_{4}$, even though the effects of unequal
quark masses $m_{3}<m_{1}<m_{2}$ have been incorporated as an
exact (without using perturbation theory) contribution to the
hamiltonian.

\subsection{Exotic baryon higher multiplets}

Let us consider exotic states ${\cal E}_{p}$ containing $p$
antiquarks and $N_{C}+p$ quarks. In the case $N_{C}=3,\,N_{f}=3$,
the only  allowed ${\cal E}_{2}$  state is a ${\bf 81}$
representation of flavor. In the particular case $N_{f}=3$, for
general $N_{C}$ the mass of the ${\cal E}_{p}$ state is
 \br M({\cal E}_{p}) = M(classical) \Big\{1 +
\frac{3}{4} (\frac{\pi}{2N_{C}})^2 \[ N_{C}(p+1)+
p(p+2)-\frac{2}{3} N_{C}^2\] + 3/2 \Big\},\label{ep}\er where the
correction is considerably larger than unity.  For example for
$N_{C}=3$ the
 mass correction becomes $3.76$ units. Even though
this correction is one unit less than the one obtained in
\cite{ellis}, we would not consider it as a consistent
semi-classical approximation for $N_{C}=3$. However, we may
consider the spacing $\D$ between ${\cal E}_{p+1}$ and ${\cal
E}_{p}$ exotic states, which for large $N_{C}$ becomes \br \D
\equiv {\cal E}_{p+1} - {\cal E}_{p} = (\frac{3}{4})
\frac{\pi^2}{4} \frac{M_{classical}}{N_{C}}\sim 3.8\, \sqrt{e_{c}
m_{q}}\,;\label{delta}\,\,\,\,\,\,\,\,\,m_{q}\equiv
\frac{2m_{1}+m_{2}}{3}\er so, the constant $\D$ of \cite{ellis} is
decreased by a factor of $3/4$. Since $M_{classical}$ is ${\cal
O}(N_{C}^{1})$, then the parameter $\D$ is a constant ${\cal
O}(N_{C}^{0})$ as the exoticity $p$ is increased. Notice that the
low-lying mesons masses are ${\cal O}(N_{C}^{0})$ in the large
$N_{C}$ limit \cite{ellis}. This would mean that the constant $\D$
value is like the addition of a meson to the $p-$state, in the
form of quark-antiquark pair, in order to progress to the next
excitation $p+1$ \cite{diakonov11}. Remember that the low-lying
mesons in the SG theory have masses $\sim 3.2\,
\sqrt{m_{q}e_{c}}$\, \cite{Rajaraman}, which are very close to the
spacing  $\D$ defined in (\ref{delta}).

\subsection{Radius parameter of the QCD$_{2}$ exotic baryons}

In QCD$_{2}$, as found above, the quantum correction to the mass
depends on one analogue of the moment of inertia appearing in four
dimensions. Following \cite{ellis} one considers \br I =
M(classical) <r^2>,\er the effective soliton radius can be defined
by \br <<r>>\equiv \sqrt{<r^2>}.\er

Let us compare the quantum mass formula (\ref{ep}) with the
corresponding relation in four dimensions \cite{Diakonov1} in the
large $N_{C}$ limit ($N_{C}>>p>>1$), so one has

\br I= \frac{8 N_{C}^2}{3 \pi^2 M_{classical}},\er and then \br <<
r>>= \sqrt{\frac{I}{M_{classical}}} = \sqrt{\frac{8}{3}}\,
\frac{N}{\pi M_{classical}}= \frac{1}{0.96 \pi N_{f}^{1/4}
\sqrt{e_{c} m_{q} }},\er where $m_{q}$ was defined in
(\ref{delta}). For $N_{f}=3$ flavors, $e_{c}= 100 MeV$ for the
coupling, and quark masses $m_{3}=4$ MeV,\, $m_{1}=54.5$ MeV,\,and
$m_{2}=55.1$MeV [these values satisfy the relationship $13m_{3}=5
m_{1}-4 m_{2}$ relevant in two-dimensions as is obtained from
(\ref{gamma2}) and (\ref{mu11})], we get for the effective baryon
radius $\approx 1/(294MeV) \sim 0.7$ fm. This is $12.5\%$ less
than the radius estimated in \cite{ellis} for QCD$_{2}$ exotic
baryons. As a curiosity, notice that the radius parameter of
$\Theta^{+}$ has been estimated to be around $1.13$ fm $= 5.65 $
GeV$^{-1}$ (see e.g. \cite{battacharya} and references therein).

\section{Discussion}

We have extended the results of refs. \cite{frishman, ellis}
concerning several properties of normal and exotic baryons by
including unequal quark mass parameters. In the case of $N_{f}=3$
flavors, the low-energy
 hadron states are described by the $su(3)$ generalized sine-Gordon model, providing a
framework for the exact computations of the lowest-order quantum
corrections of various  quantities, such as the masses of the
 normal and exotic baryons. The semi-classical quantization method we adopted
is an  analogue of the rigid-rotor approach (RRA) applied in four
dimensional QCD to quantize normal and exotic baryons (see e.g.
\cite{Diakonov1}). Even though there is no spin in 2D, we have
compared our results to their analogues in 4D; so, obtaining
various similarities to the results from
 the chiral-soliton approaches in QCD$_{4}$. The RRA we have followed, as discussed in section \ref{exo}, may be
justified in our case since there is no mixing between the
intrinsic vibrational modes and the collective rotation in flavor
space degrees of freedom \cite{cohen}. It is remarkable that the
GSG ansatz (\ref{diag0}),  with soliton solutions which take into
account the unequal quark mass parameters, allowed us to improve
the lowest order quantum corrections for various physical
quantities, such as the baryon masses; in this way rendering the
semi-classical method more reliable in the large $N_{C}$ limit.

Other properties of the baryons such as a proper treatment of
$k-$baryon bound states (extending the results of
\cite{frishman90} for GSG type ansatz), including baryon-meson
scattering amplitudes, are still to be addressed in the future.

We have found that the remarkable double sine Gordon model arises
as a reduced GSG model bearing a kink($K$) type solution
describing a multi-baryon; so, the description of some resonances
in QCD$_{2}$ may take advantage of the properties of the
$K\bar{K}$ system which are being considered in the current
literature \cite{mussardo, mussardo1, campbell}.

\vskip 0.2in

{\sl Acknowledgements}

The author thanks H.L. Carrion for collaboration in a previous
work and the Mathematics and Physics Departments-UFMT (Cuiab\'a)
and IMPA (Rio de Janeiro) for the kind hospitality. This work has
been supported by CNPq-FAPEMAT.

\appendix

\section{$sl(3, \IC)$ GSG model, soliton and kink solutions}

Here we summarize some properties of the $sl(3, \IC)$ GSG model
\cite{jhep4, epjc} relevant to our discussions above, such as the
soliton and kink spectrum. The discussions make some connection to
the QCD$_{2}$ developments above, such as (multi-) baryon number
of solitons and kinks. The third soliton solution with baryon
number $N_{C}$ is new.

The generalized sine-Gordon model (GSG) related to $sl(3, \IC)$ is
defined by \cite{jmp, jhep, epjc} \br \label{GSG1} S= \int d^2x
\sum_{i=1}^{3}\[ \frac{1}{2} (\pa_{\mu} \Phi_{i})^2 + \mu_{i} \(
\mbox{cos} \b_{0} \Phi_{i}-1\)\]. \er

Since in this case one has two simple roots there are two
independent real fields, $\vp_{1, \,2}$, such that \br
\label{fields} \Phi_{1}= \nu_{1}(2\vp_{1}-\vp_{2});\,\,\,\Phi_{2}=
\nu_{2} ( 2\vp_{2}-\vp_{1});\,\,\,\Phi_{3}= \nu_{3}(r\, \vp_{1}+
s\, \vp_{2}),\\\nonumber \nu_{i}\,(i=1,2,3),\,s,\,r \in \IR \er
which must satisfy the constraint
\begin{eqnarray} \label{constr}
\Phi_{3}= \d_{1}  \Phi_{1}+\d_{2}  \Phi_{2},
\end{eqnarray}
where $\d_{1}, \d_{2}$ are some real numbers. The $\Phi$ fields
dependence on the $\vp'$s
 can be explained in the context of the Lie algebraic
construction of the classical version of the model \cite{jhep,
matter}.

Taking into account (\ref{fields})-(\ref{constr}) and the fact
that the fields $\vp_{1}$ and $\vp_{2}$ are independent we may get
the relationships \br\label{nus} \nu_{2} \d_{2} = \rho_{0} \nu_{1}
\d_{1} \,\,\,\,\,\nu_{3} =\frac{1}{r+s}(\nu_{1} \d_{1}+\nu_{2}
\d_{2} );\,\,\,\, \rho_{0} \equiv \frac{2s+r}{2r+s} \er

The $sl(3, \IC)$ model has a potential density \br V[\vp_{i}] =
\sum_{i=1}^{3} \mu_{i}\(1- \mbox{cos} \b_{i}\Phi_{i}\)
\label{potential}\er

In the $sl(3, \IC)$ construction of \cite{epjc} the parameters
$\d_{i}$ depend on the couplings $\b_{i}\, [\b_{i}\equiv \b_{0}
\nu_{i}]$ and they satisfy certain relationship. This is obtained
by assuming $\mu_{i}>0$ and the zero of the potential given for
$\Phi_{i}= \frac{2\pi}{\b_{0}} n_{i}$, which substituted into
(\ref{constr}) provides \br \label{deltas} n_{1} \d_{1}+ n_{2}
\d_{2}= n_{3},\,\,\,\,n_{i} \in \IZ \er

The last relation combined with (\ref{nus}) gives \br (2r+s)
\frac{n_{1}}{\nu_{1}}+ (2s+r) \, \frac{n_{2}}{\nu_{2}} = 3
\,\frac{n_{3}}{\nu_{3}}\label{nns}.\er

The periodicity of the potential implies an infinitely degenerate
ground state and then the theory supports topologically charged
excitations. So, consider the vacuum lattice defined by \br
(\Phi_{1}\,, \, \Phi_{2}) =\frac{2\pi}{\b_{0}} (n_{1}
\,,\,n_{2}),\,\,\,\,\,n_{a} \in \IZ. \label{lattice1}\er

It is convenient to write the equations of motion in terms of the
independent fields $\vp_{1}$ and $\vp_{2}$  \br \pa^2 \vp_{1} &=&
- \mu_{1} \b_{1} \D_{11} \mbox{sin} [\b_{1} ( 2\vp_{1}-\vp_{2})]-
\mu_{2} \b_{2} \D_{12} \mbox{sin}[\b_{2} (2\vp_{2}-\vp_{1})]+
\nonumber\\&& \mu_{3} \b_{3} \D_{13} \mbox{sin} [\b_{3} ( r
\vp_{1} + s \vp_{2})]
\label{eq1}\\
\pa^2 \vp_{2} & = & - \mu_{1} \b_{1} \D_{21} \mbox{sin} [\b_{1} (
2\vp_{1}-\vp_{2})]-\mu_{2} \b_{2} \D_{22} \mbox{sin}[\b_{2}
(2\vp_{2}-\vp_{1})]+ \nonumber\\  && \mu_{3} \b_{3} \D_{23}
\mbox{sin} [\b_{3} ( r \vp_{1}+ s \vp_{2})]\label{eq2}, \er where
the $\D_{ij}'s$ depend on $\b_{0},\,\nu_{j}\,(j=1,2,3),
\,r,\,s,\,\d_{a}\,(a=1,2)$.

Notice that the eqs. of motion (\ref{eq1})-(\ref{eq2}) exhibit the
symmetries \br\label{symm} \vp_{1} &\leftrightarrow &
\vp_{2},\,\,\,\,\mu_{1} \leftrightarrow \mu_{2}, \,\,\,\nu_{1}
\leftrightarrow \nu_{2},\,\,\,\d_{1} \leftrightarrow \d_{2},\,\,\,
r \leftrightarrow s; \\
&\mbox{and}&\,\,\,\vp_{a} \leftrightarrow - \vp_{a},\,\,\,a=1,2
\label{changesign}\er

In the following we write the 1-soliton(antisoliton),
1-kink(antikink) and bounce type solutions and compute the
relevant (multi-)baryon numbers associated to the $U(1)$ symmetry
in the context of QCD$_{2}$.

\subsection{One soliton/antisoliton pair associated to $\vp_{1}$}
\label{s11}

The functions  \br \label{sol1} \vp_{1}=
\frac{4}{\b_{0}}\mbox{arctan}\{d\,\, \mbox{exp}[\gamma_{1} (x-v
t)]\},\,\,\,\,\vp_{2}=0, \er satisfy the system of equations
(\ref{eq1})-(\ref{eq2}) for the set of parameters \br
\label{para1} \nu_{1}=1/2,\,\,
\delta_{1}=2,\,\,\delta_{2}=1,\,\,\nu_{2}=1,\,\,
\nu_{3}=1,\,\,r=1.\er provided that \br \label{gamma1}
13\mu_{3}=5\mu_{2}-4\mu_{1},\,\,\,\,
\gamma^2_{1}=\frac{\b_{0}^2}{13}(6\mu_{2}+3\mu_{1}).\er

This solution is precisely the sine-Gordon 1-soliton associated to
the field $\vp_{1}$ with mass \br M_{1}^{sol}=\frac{8
\g_{1}}{\b^2_{0}}.\label{solmass1}\er

From (\ref{fields}) and taking into account the parameters
(\ref{para1}) one has the relationships between the GSG fields \br
\Phi_{1} = -\Phi_{2} =  \Phi_{3} = \, \vp_{1} \label{relation1}\er

Moreover, from (\ref{deltas})-(\ref{nns}) and (\ref{para1}) one
gets the relationships \br n_{1}=-n_{2}=n_{3} \label{nn1}\er

Taking into account the QCD$_{2}$ motivated formula (\ref{baryon})
 and (\ref{nn1}) one can compute the baryon number of the GSG
 soliton (\ref{sol1}) taking $n_{1}=1$
 \br
 \label{charge1}
 {\cal Q}_{B}^{(1)} = N_{C},
 \er
where the superindex $(1)$ refers to the associated $\vp_{1}$
field nontrivial solution.

\subsection{One soliton/antisoliton pair associated to $\vp_{2}$}
\label{s22}

The functions \br \label{sol2} \vp_{2}= \frac{4}{\b_{0}}
\mbox{arctan}\{d\,\mbox{exp}[\gamma_{2}(x-vt)]\},\,\,\,\,\vp_{1}=0\er
solve the system (\ref{eq1})-(\ref{eq2}) for the choice of
parameters \br \label{para2} \nu_{1}=1,\,\,
\delta_{1}=1,\,\,\delta_{2}=2,\,\,\nu_{2}=1/2,\,\,
\nu_{3}=1,\,\,s=1\er provided that \br
13\mu_{3}=5\mu_{1}-4\mu_{2},
\,\,\,\,\gamma^2_{2}=\frac{\b_{0}^2}{13}(6\mu_{1}+3\mu_{2})
\label{gamma2}. \er

This is the sine-Gordon 1-soliton associated to the field
$\vp_{2}$ with mass \br M_{2}^{sol} = \frac{8
\g_{2}}{\b^2_{0}}.\label{solmass2}\er

As above from (\ref{fields}) and the set of parameters
(\ref{para2}) one has the relationships \br - \Phi_{1} = \Phi_{2}
=  \Phi_{3} =  \vp_{2}.\label{relation2}\er

From (\ref{deltas})-(\ref{nns}) and (\ref{para2}) one gets the
relationship \br n_{1}=-n_{2}=-n_{3} \label{nn2}.\er

So, taking into account the QCD$_{2}$ motivated formula
(\ref{baryon}) and (\ref{nn2})
 one  computes the baryon number of this GSG soliton taking $n_{2}=1$
 \br
 \label{charge2}
 {\cal Q}_{B}^{(2)} = N_{C},
 \er
where the superindex $(2)$ refers to the associated $\vp_{2}$
field.

\subsection{1-soliton/1-antisoliton pairs associated to $\hat{\vp} \equiv \vp_{1}=\vp_{2}$}
\label{s33}

In the case $\vp_{1}=\vp_{2}$ one has the 1-soliton solution
$\hat{\vp}$  of the system (\ref{eq1})-(\ref{eq2})
 associated to the parameters
 \br \label{para3} \nu_{1}=1,\,
\,\delta_{1}=-1/2,\,\,\nu_{2}=1,\,\,
 \delta_{2}=-1/2,\,\, \nu_{3}=-1/2,\,\,r=s=1.\er

One has the 1-soliton \br \vp_{1}&=&\vp_{2}\equiv
\hat{\vp},\,\,\,\, \nonumber
\\\hat{\vp} &=& \frac{4}{\b_{0}} \mbox{arctan}\{d\,\,
\,\mbox{exp}[\gamma_{3}(x-vt)]\}, \label{sol3b}\er which requires
\br\label{gamma3} \gamma^2_{3} = \b_{0}^2
\(\mu_{1}+\frac{1}{2}\mu_{3}\),\,\,\,\,\,\mu_{1}=\mu_{2}. \er

This is a sine-Gordon 1-soliton associated to both fields
$\vp_{1,\,2}$ in the particular case when they are equal to each
other.  It possesses a mass \br
M_{3}^{sol}=\frac{8\g_{3}}{\b_{0}^2}.\label{solmass3}\er

In view of the symmetry (\ref{symm}) which are satisfied by the
parameters (\ref{para3}) and (\ref{gamma3}) one can think of this
solution as doubly degenerated.

As above, from (\ref{fields}) and the set of parameters
(\ref{para3}) one has the following relationships  \br  \Phi_{1} =
 \Phi_{2} =  - \Phi_{3} =
\hat{\vp} .\label{relation3}\er

From (\ref{deltas})-(\ref{nns}) and (\ref{para3}) one gets the
relationship \br -2n_{3}=n_{1}+ n_{2} \label{nn3}.\er So, taking
into account the QCD$_{2}$ motivated formula (\ref{baryon}) and
(\ref{nn3})
 one  computes the baryon number of this GSG solution taking $n_{3}=-1$
 \br\label{charge3}
 {\cal Q}_{B}^{(\hat{\vp})} = N_{C}, \er
where the superindex refers to the associated $\hat{\vp}$ field.

\subsubsection{Antisolitons and general N-solitons}

The GSG system (\ref{eq1})-(\ref{eq2}) reduces to the usual SG
equation for each choice of the parameters (\ref{para1}),
(\ref{para2}) and (\ref{para3}), respectively. Then, the
$N-$soliton solutions in each case can be constructed as in the
ordinary sine-Gordon model.

Using the symmetry (\ref{changesign}) one can be able to construct
the {\sl 1-antisolitons} corresponding to the soliton solutions
(\ref{sol1}), (\ref{sol2}) and (\ref{sol3b}) simply by changing
their signs $\vp_{a} \rightarrow -\vp_{a}$.

\subsection{Mass splitting of solitons}
\label{splitt}

It is interesting to write some relationships among  the various
soliton masses.

i) For $\mu_{1}\neq \mu_{2}$ one has respectively the two
1-solitons, (\ref{sol1}) and (\ref{sol2}), with masses
(\ref{solmass1}) and (\ref{solmass2}) related by \br
(M_{1}^{sol})^2 - (M_{2}^{sol})^2= \frac{48N_{C}}{\pi}
(\mu_{2}-\mu_{1}).\label{relat12} \er

ii) For $\mu_{1}=\mu_{2}$, there appears the third soliton
solution (\ref{sol3b})-(\ref{gamma3}). Then, taking into account
(\ref{gamma1}), (\ref{gamma2}), (\ref{gamma3}), (\ref{relat12})
and the third soliton mass (\ref{solmass3}) we have the
relationships \br \label{relat123}
M_{1}^{sol}&=&M_{2}^{sol},\,\,\,\,\,\,\,M_{3}^{sol} = \sqrt{3/2}\,
M_{1}^{sol},\\ \label{gammas} \g_{1} &=& \g_{2} = \sqrt{2/3}\,\,
\g_{3},\,\,\,\,\,\mu_{3}=\frac{1}{13} \, \mu_{1}. \er

Notice that in this case  $M_{3}^{sol} < M_{1}^{sol}+
M_{2}^{sol}$, and the third soliton is stable in the sense that
energy is required to dissociate it.

\subsection{Kink of the double sine-Gordon model as a multi-baryon}
\label{dsg:sec}

In the system (\ref{eq1})-(\ref{eq2}) we perform the following
reduction $\phi \equiv \vp_{1}=\vp_{2}$ such that \br
\Phi_{1}=\Phi_{2},\,\,\,\Phi_{3}= q \, \Phi_{1}, \label{reduc}\er
with $q$ being a real number.

Moreover, for consistency of the system of equations
(\ref{eq1})-(\ref{eq2}) we have
 \br \mu_{1}=\mu_{2},\,\,\,\,\d_{1} =\d_{2}=q/2
,\,\,\,\nu_{1}=\nu_{2},\,\,\,\nu_{3}=\frac{q}{2}
\nu_{1},\,\,r=s=1. \label{param1}\er

Thus the system of Eqs.(\ref{eq1})-(\ref{eq2}) reduces to \br
\label{dsg}\pa^2 \Phi_{DSG} &=& -
\frac{\mu_{1}}{\nu_{1}}\,\mbox{sin}(\nu_{1}
\Phi_{DSG})-\frac{\mu_{3} \d_{1}}{\nu_{1}} \mbox{sin} (q\, \nu_{1}
\Phi_{DSG}),\,\,\,\,\, \Phi_{DSG} \equiv \b_{0} \phi. \er

This is the so-called {\sl two-frequency  sine-Gordon} model (DSG)
and it has been the subject of much interest in the last decades,
from the mathematical and physical points of view.

If the parameter $q$ satisfies \br \label{frac1} q =  \frac{n}{m}
\, \in \,  \IQ \er with $m, \,n$ being two relative prime positive
integers, then the potential
$\frac{\mu_{1}}{\nu_{1}^2}(1-\mbox{cos} (\nu_{1} \Phi_{DSG}))
+\frac{\mu_{3}}{2 \nu_{1}^2}(1-\mbox{cos} (q \nu_{1} \Phi_{DSG}))$
associated to the model (\ref{dsg}) is periodic with period \br
\frac{2\pi}{\nu_{1} } m = \frac{2\pi}{q\, \nu_{1} } n. \er

Then, as mentioned above the theory (\ref{dsg}) possesses
topological excitations.

From (\ref{fields}) and the set of parameters (\ref{param1}) one
has the relationships \br   \Phi_{1} = \Phi_{2} = \frac{1}{q}
\Phi_{3} = \nu_{1} \phi .\label{fieldskk}\er

And from (\ref{deltas})-(\ref{nns}) and (\ref{param1}) one gets
the relationship \br n_{3}=\frac{q}{2}(n_{1}+ n_{2})
\label{nn4}.\er

So, taking into account the QCD$_{2}$ motivated formula
(\ref{baryon}) and (\ref{nn4})
 one  computes the baryon number of this DSG solution
 \br\label{bar3}
 {\cal Q}_{B}^{(DSG)} =  N_{C} (1+\frac{2}{q}) n_{3}, \,\,\,\,n_{3} \in \IZ, \er
where the superindex $(DSG)$ refers to the associated DSG
solution.

In the following we will provide some kink solutions for a
particular set of parameters. Consider \br
\label{paras}\nu_{1}=1/2,\,\,
\d_{1}=\d_{2}=1,\,\,\nu_{2}=1/2,\,\,\,\nu_{3}=1/2 \,\,\,
\mbox{and} \,\,\, q=2,\,n=2,\, m=1\er which satisfy (\ref{param1})
and (\ref{frac1}). This set of parameters provide the so-called
{\sl double sine-Gordon model} (DSG), such that from
(\ref{fieldskk}) and (\ref{paras}) the field configurations
satisfy \br \Phi_{1} = \Phi_{2} = \frac{1}{2} \Phi_{3} =
\frac{1}{2} \phi .\label{fieldskk1}\er

Its potential $-[4 \mu_{1}(\mbox{cos} \frac{\Phi_{DSG}}{2}-1 )+
2\mu_{3}(\mbox{cos} \Phi_{DSG} -1)]$ has period $4\pi$ and has
extrema at $\Phi_{DSG} = 2\pi p_{1}$, and\, $\Phi_{DSG} = 4\pi
p_{2} \pm 2 \mbox{cos}^{-1} [1-|\mu_{1}/(2\mu_{3})|]$ with
$p_{1},p_{2} \in \IZ$; the second extrema exists only if
$|\mu_{1}/(2\mu_{3})|< 1$. Depending on the
 values of the parameters $\b_{0},\, \mu_{1},\,\mu_{3}$ the quantum field theory
 version of the DSG model presents a variety of physical effects, such as the decay
 of the false vacuum, a phase transition, confinement of the kinks and the resonance
 phenomenon due to unstable bound states  of excited kink-antikink states (see \cite{mussardo} and
 references therein). The
semi-classical spectrum of neutral particles in the DSG theory is
investigated in \cite{mussardo1}.

A particular solution of (\ref{dsg}) for the parameters
(\ref{paras}) can be written as \br \label{gen} \Phi_{DSG} := 4\,
\mbox{arctan}\left[\frac{1}{d}\,\,\frac{1+h\,\,\mbox{exp}[2\gamma(x-vt)]}{\mbox{exp}[\gamma(x-vt)]}\right]
\er provided that \br \label{gammak}
\gamma^2&=&\b_{0}^2\(\mu_{1}+2\mu_{3}\),\,\,\,\,\,\, h=
-\frac{\mu_{1}}{4},\,\,\,\,\label{h1}\er

\subsubsection{A multi-baryon and the DSG kink ($h < 0, \,\mu_{i}>0$)}

For the choice of parameters $h < 0, \,\mu_{i}>0 $ in (\ref{h1})
the equation (\ref{gen}) provides \br \label{kink} \phi
&=&\frac{4}{\b_{0}} \mbox{arctan}\left[\frac{-2
|h|^{1/2}}{d}\,\,\mbox{sinh}[\gamma_{K}\, (x-vt) +
a_{0}]\right],\,\,\,\,\g_{K} \equiv \pm \b_{0} \sqrt{\mu_{1}+ 2
\mu_{3}},\\
\nonumber && a_{0}=\frac{1}{2} \mbox{ln} |h|. \er

This is the DSG 1-kink solution with mass \br \label{kinkmass}
M_{K} = \frac{16}{\b_{0}^2} \g_{K} \left[1
+\frac{\mu_{1}}{\sqrt{2\mu_{3} (\mu_{1}+ 2\mu_{3})}}\mbox{ln}
(\frac{\sqrt{\mu_{1}+ 2\mu_{3}}+
\sqrt{2\mu_{3}}}{\sqrt{\mu_{1}}})\right]. \er

Since one must have $\frac{\mu_{3}}{\mu_{1}}>\frac{1}{2}$ (see
below for the range of possible values of these parameters) the
potential supports one type of minima and thus there exists only
one type of topological kink \cite{campbell}. So, the DSG model
possesses
 only the topological excitation (\ref{kink}) relevant to our QCD$_{2}$
discussion.

One can relate the parameters $\mu_{j}$ in (\ref{GSG1}) to the
mass parameters $m_{i}$ in the effective lagrangian of QCD$_{2}$
 in (\ref{GSG}). So, for the ``physical values" \, $N_{f} = 3$\, and $e_{c}$ = 100MeV for the
coupling and taking into account (\ref{mass}), (\ref{mlargen}) and
(\ref{masspar}) one has for large $N_{C}$ \br \mu_{j} = 2
\frac{m_{j}}{m_{0}} m^2\, \approx\, N_{C} \,m_{j}\,
124(\mbox{MeV}), \label{mu11} \er thus, the $\mu_{j}'s$ have
dimension (MeV)$^2$.

For the values of the mass parameters $\mu_{1}, \,\mu_{3}$ in the
 range $[10^{3}\,,\,5 \times 10^{4}]$(MeV)$^2$\, (take $m_{1}\approx m_{2} \approx 52$ MeV;\, $m_{3}=4$ MeV, notice that these values
 satisfy the relationship (\ref{gamma2})) one can determine the values of the ratio $\kappa$
 between the kink (\ref{kinkmass}) and the third soliton (\ref{solmass3}) masses
 \br \kappa \equiv
\frac{M_{K}}{M_{3}^{sol}},\,\,\,\,\,\,\,\, 4<\kappa <
4.2\label{ratio}\er

The baryon number of this DSG kink solution is obtained from
(\ref{bar3}) taking $q=2,\,\,n_{3}=2$
 \br
 {\cal Q}_{B}^{(K)} =4N_{C}, \label{chargekink}\er
where the superindex $(K)$ refers to the associated DSG kink
solution.

The above relations (\ref{ratio})-(\ref{chargekink}) suggest that
the decay of the {\sl kink} to four solitons
$\{M^{sol}_{j}\}\,(j=1,2,3)$ is allowed by conservation of energy
and charge, however one can see from the kink dynamics that it is
a stable object and its fission may require an external trigger.
For similar phenomena in soliton dynamics see ref. \cite{riazi}.

Let us emphasize that the baryons with charges\, $2n_{3}N_{C}$\, [
set $q=2$\, in (\ref{bar3})] for $n_{3}=1,2,...$ are assumed to be
bound states of $2, 4,...$ ``basic" baryons, and so, they would
correspond to di-baryon states like {\sl deuteron}
($\,^{1}_{1}H^{+}$) and
 the ``$\a$ particle" ($\,^{4}_{2}He^{+}$). However, we have not
 found, for the QCD$_{2}$ motivated parameter space $(\mu_{1}, \mu_{3})$\, any
  kink with baryon number $2N_{C}$. These $2-$baryons are expected
  to be found in the $2-$soliton sectors of the GSG model. Notice that in our formalism the four-baryon
  appears already for
$N_{f}=3$ as a DSG kink with topological charge
(\ref{chargekink}). In the formalism of refs. \cite{frishman,
frishman90} the multibaryons have baryon number $k N_{C}$\, ($k
\leq N_{f}-1$), so the $(N_{f}-1)-$baryon is the one with the
greatest baryon number.

\subsection{Configuration with baryon number $3N_{C}$}

These solutions do not form stable configurations, nevertheless we
 describe them for completeness. Let us take $\vp_{1}=\vp_{2}$, so
one has two 1-soliton solutions $\hat{\vp}_{A}\,(A=1,2)$  of the
system (\ref{eq1})-(\ref{eq2})
 associated to the parameters
 \br \label{para33} \nu_{1}=1,\,
\,\delta_{1}=1/2,\,\,\nu_{2}=1,\,\,
 \delta_{2}=1/2,\,\, \nu_{3}=1/2,\,\,r=s=1.\er

As the first 1-soliton one has \br \vp_{1}&=&\vp_{2}\equiv
\hat{\vp}_{1},\,\,\,\, \label{sol33a}
\\\hat{\vp}_{1} &=& \frac{4}{\b_{0}} \mbox{arctan}\{d\,\,
\,\mbox{exp}[\gamma_{4}(x-vt)]\}, \label{sol33b}\er which requires
\br\label{gamma33} d^2=1, \,\,\,\,38\gamma^2_{4} =\b_{0}^2\(
25\mu_{1}+13\mu_{2}+19\mu_{3}\) \er

This is a sine-Gordon 1-soliton associated to both fields
$\vp_{1,\,2}$ in the particular case when they are equal to each
other.  It possesses a mass \br
 M_{4}^{sol}=\frac{8\g_{4}}{\b_{0}^2}.\label{solmass4}\er

In view of the symmetry (\ref{symm}) we are able to write from
(\ref{gamma33}) \br d^2=1, \,\,\,\,38\gamma^2_{5} =
25\mu_{2}+13\mu_{1}+19\mu_{3}, \er and then one has another
1-soliton  from (\ref{sol33a})-(\ref{sol33b}) \br
\vp_{1}&=&\vp_{2}\equiv \hat{\vp}_{2},\, \label{sol33c}
\\\hat{\vp}_{2}&=& \frac{4}{\b_{0}} \mbox{arctan}\{d\,\, \,\mbox{exp}[\gamma_{5}(x-vt)]\}.
\label{sol33d}\er

It possesses a mass \br
M_{5}^{sol}=\frac{8\g_{5}}{\b_{0}^2}.\label{solmass5}\er

Similarly, from (\ref{fields}) and the set of parameters
(\ref{para33}) one has the following relationships  \br  \Phi_{1}
=
 \Phi_{2} =  \Phi_{3} =
\hat{\vp}_{A},\,\,\,\,A=1,2 .\label{fields333}\er

From (\ref{deltas})-(\ref{nns}) and (\ref{para33}) one gets the
relationship \br 2n_{3}=n_{1}+ n_{2} \label{nn33}.\er

So, taking into account the QCD$_{2}$ motivated formula
(\ref{baryon}) and (\ref{nn3})
 one  computes the baryon number of this GSG solution taking $n_{3}=1$
 \br
 \label{topotri}
 {\cal Q}_{B}^{(A)} = 3N_{C}, \er
where the superindex $(A)$ refers to the associated
$\hat{\vp}_{A}$ field. Therefore, the both solutions $A=1,\,2$,
have the same baryon number in the context of QCD$_{2}$. The
 individual soliton solutions (\ref{sol33b}) and (\ref{sol33d})
 have, each one, a topological charge $N_{C}$, since they are sine-Gordon
 solitons. Then, the configuration (\ref{fields333}) with total charge $3N_{C}$ is
 composed of three SG solitons. Therefore, by
 conservation of energy and topological charge arguments one has that
the rest mass of the static configurations \, $A=1,2$,\, with
baryon number $3N_{C}$  will be, respectively \br
M_{4,\,5}^{config.} \equiv 3\,M_{4,\, 5}^{sol}\label{masstri},\er
where the masses $M_{4,\, 5}^{sol}$ are given by (\ref{solmass4}),
(\ref{solmass5}).

Moreover, one can verify the following relationships
\br\label{ineq1} i)\,\, M_{4,\,5}^{config.} & >& M_{1}^{sol} +
M_{2}^{sol};\,\,\,\,\,\,\,\,\,\,\,\,\,\,\,\,\,\,\,\,\,\,\,\,\,\,\,\,\,\,\,\,
\,\,\,\,\,\,\,\,\,\,\,\,\,\,\,\,\,\,\,\,\,\,\,\,\,\,\,\,\,\,\,\, \mu_{1} \neq \mu_{2},\\
ii) \,\,M_{4}^{config.}&=&M_{5}^{config.}
> M_{1}^{sol} + M_{2}^{sol}+
M_{3}^{sol};\,\,\,\,\,\,\,\,\,\,\,\,\,\mu_{1} =
\mu_{2},\label{ineq2}\er where the soliton masses
$M_{j}^{sol}\,\,(j=1,2,3)$ are given by (\ref{solmass1}),
(\ref{solmass2}), (\ref{solmass3}), respectively. One observes
that the configurations $A=1,2$,\, do not form  bound states
(bound states would be formed if the inequalities
(\ref{ineq1})-(\ref{ineq2})
 are reversed), and they may decay into the ``basic" set
$\{M_{1}^{sol},\, M_{2}^{sol}\}$\,\,or\,\, $ \{M_{1}^{sol},\,
M_{2}^{sol},\, M_{3}^{sol}\}$\, of solitons, such that the excess
energy is transferred to the kinetic energy of the solitons.

\end{document}